# Synthesis and structure of carbon doped $H_3S$ compounds at high pressure

Alexander F. Goncharov[1], Elena Bykova[1], Maxim Bykov[1,2], Xiao Zhang[3], Yu Wang[3], Stella Chariton[4], Vitali B. Prakapenka[4],

[1] *Earth and Planets Laboratory, Carnegie Institution of Washington, 5251 Broad Branch Road NW, Washington, DC 20015, USA*

[2] *Department of Mathematics, Howard University, Washington, DC 20059, USA*

[3] *Key Laboratory of Materials Physics, Institute of Solid State Physics, HFIPS, Chinese Academy of Sciences, Hefei 230031, Anhui, People's Republic of China*

[4] *Center for Advanced Radiations Sources, University of Chicago, Chicago, Illinois 60637, USA*

**Understanding of recently reported putative close-to-room-temperature superconductivity in C-S-H compounds at 267 GPa demands reproducible synthesis protocol as well as knowledge of its structure and composition. We synthesized C-S-H compounds with various carbon composition at high pressures from elemental C and methane $CH_4$, sulfur S, and molecular hydrogen $H_2$. Here we focus on compounds synthesized using methane as these allow a straightforward determination of their structure and composition by combining single-crystal X-ray diffraction (XRD) and Raman spectroscopy. We applied a two-stage synthesis of $((CH_4)_x(H_2S)_{(1-x)})_2H_2$ compounds by first reacting sulfur and mixed methane-hydrogen fluids and forming $CH_4$ doped $H_2S$ crystals at 0.5-3 GPa, and then by growing single crystals of the desired hydrogen rich compound. Raman spectroscopy applied to this material shows the presence of the $CH_4$ molecules incorporated into the lattice and allows to determine the $CH_4$ content, while single-crystal X-ray diffraction results suggest that the methane molecules substitute $H_2S$ ones. The structural behavior of these compounds is very similar to the previously investigated methane free compounds demonstrating a transition from $Al_2Cu$ type *I*4/*mcm* structure to a modulated structure at 20-30 GPa and back to the same basic structure in an extended modification with greatly modified Raman spectra. This latter phase demonstrates a distortion into *Pnma* structure at 132-159 GPa and then transforms into a common *Im*-3*m* $H_3S$ phase at higher pressures, however, no structural anomaly is detected near 220 GPa, where a sharp upturn in $T_c$ has been reported.**

Study of high-pressure stabilized poly- and super-hydrides continues to be of great interest because of very high superconducting transition temperatures $T_c$ reported experimentally and theoretically [1-6]. The record-high close-to-room temperature superconductivity are achievable due to very high phonon frequencies related to vibrations of light hydrogen atoms and a strong electron-phonon coupling, which are the necessary factors affecting $T_c$ in the Bardeen-Cooper-Schrieffer (BCS) theory. However, $T_c$ critically depends on the atomic and electronic structures of the material, which are in turn dependent on the chemical composition and bonding types and their geometry. Finding the structure and composition of candidate superconducting compounds with the optimal properties presents a great challenge because of their critical variability depending on composition and pressure. Experiments at high pressure are very challenging in determination the structure and composition especially for hydrogen bearing compounds and these challenges are greatly increased at pressures of 140-270 GPa, where superconductivity was reported.

Sulfur polyhydride, $H_3S$, demonstrated a very high $T_c$ of 203 K at 150 GPa [1]. This compound forms a cubic *Im*-3*m* structure with covalent S-H bonds. Recently, Sneider *et al.*[4] reported even higher values of $T_c$ in a mixed C/S polyhydride compressed up to 267 GPa, but the composition and structure of this materials has not been determined. It has been proposed that the synthesized at 4 GPa compound has *I*4/*mcm* symmetry similar to an extensively studied $(H_2S)_2H_2$ and $((CH_4)_x(H_2S)_{(1-x)})_2H_2$ composition with x≈0.5 [7-11]. Theoretical calculations show that this phase is a narrow gap semiconductor [11], and, thus, even if it becomes metallic under pressure, is not expected to possess high $T_c$ because of a pseudogap at the Fermi level. In contrast, theoretical calculations on this system in *Im*-3*m* structure [12, 13] demonstrated high-temperature superconductivity comparable to pure $H_3S$ but no sharp upturn of $T_c$ above 220 GPa as reported by Sneider *et al.* [4] is predicted. However, the most recent theoretical calculations suggest that rather small C doping (4-6%) can result in a substantial boosting of $T_c$ in *Im*-3*m* $H_3S$ [14, 15], but this result has been questioned recently [16]. Moreover, carbon doping can result in a structural change and recent experiments showed the existence of an orthorhombic structure in the pressure range above 140 GPa, where cubic *Im*-3*m* was expected to be stable [17, 18]. These experimental and theoretical results underscore the importance of knowledge of the structure and composition of superconducting materials for understanding the mechanism of high-temperature superconductivity.

The synthesis of carbonaceous sulfur hydride at 4 GPa by Sneider *et al.* [4] was reportedly via photochemistry from elemental components, however the mechanism of this process and, most importantly, the pathway that creates methane, one of the assumed building blocks, was not described. Moreover, the carbon composition of the compounds synthesized could be neither controlled nor determined. A similar "photochemical" synthesis of the carbon-free $(H_2S)_2H_2$ compound has been previously reported [19], likely inspiring Sneider *et al.* [4] to perform their synthesis of carbonaceous material. In contrast, Bykova *et al*. work [17] was not able to reproduce the reported synthesis procedure from a mixture of elemental C and S in that they find no sign of carbon that can be detected by X-ray diffraction (XRD) and Raman spectroscopy in the $(H_2S)_2H_2$ crystals grown at 4 GPa using the laser assisted procedure similar to that described by Sneider *et al.* [4]. Since $(H_2S)_2H_2$ compound can be synthesized by a handful of different techniques [7-10] including spontaneous crystallization from $H_2S$ and $H_2$ [8], other synthesis pathways can be explored. Here we establish a new synthesize route for $((CH_4)_x(H_2S)_{(1-x)})_2H_2$ compounds with a variable x (0.01<x<0.1) by reacting $H_2$-$CH_4$ gas mixture of various $CH_4$ composition with elemental sulfur. Based on the Raman experiments, we show that the synthesized compound incorporates $CH_4$ proportionally to the methane composition in the gas mixture. The synthesized here materials with x≈0.01-0.03 have virtually the same vibrational properties under pressure as those reported by Sneider *et al*. [4] suggesting that our single-crystal XRD measurements performed here to 240 GPa are relevant to the reported superconductivity. We find the same phase sequence *I*4/*mcm*→*Pnma*→*Im*-3*m* under pressure as previously found by our group for the compound with the estimated composition of x≈0.5 [17], which appears to be near 0.07(1) based on the analysis presented in this work.

We synthesized $((CH_4)_x(H_2S)_{(1-x)})_2H_2$ samples in a BX-90 diamond anvil cell[20] (DAC) equipped with Boehler-Almax diamond anvils [21] with 200 μm in diameter flat culet and beveled and toroidal shaped using FIB culet [22] with a central flat culet of 40 μm in diameter for experiments to up to 64 and 250 GPa, respectively (Table 1). A small piece of crystalline sulfur was positioned on an anvil, gold and ruby served as pressure gauges for XRD and optical experiments, respectively. Methane and hydrogen were loaded in the DAC cavity sequentially at room temperature with a variable partial pressures of 1:50 to 1:1 using a compressor to a maximum pressure of 0.16 GPa (Table 1). The gases were allowed to mix for 30-40 minutes in the high-pressure vessel before the DAC was clamped. Raman spectroscopy confirmed that methane and hydrogen largely mix after 30-40 minutes equilibration time demonstrating a monotonic variation of the Raman C-H and H-H stretching modes intensity ratio depending on the initial partial pressures (Fig. S1 of Supplementary Material [23]). By applying van der Waals equations of state of $CH_4$ and $H_2$ and using the experimental gas partial pressures and gas loading sequence we calculated the gas composition in all the experiments (Table 1). In addition, Raman spectroscopy has been employed to determine the $CH_4$ composition in two high-pressure regimes: fluid $CH_4$-$H_2$ mixture after gas loading at 0.5-5 GPa (Fig. S1 of Supplementary Material [23]) and $((CH_4)_x(H_2S)_{(1-x)})_2H_2$ crystals (Fig. 1(a)) after synthesis at 4.5-7 GPa. The $CH_4$ content was determined from the peak area ratios assuming the peak intensity calibration performed up to 30 MPa [24] holds at high pressure. This assumption likely works well in the mixed fluid state, while it requires a confirmation in the solid state, where the Raman intensities can be affected by intermolecular coupling. In one experiment (XZ003), a commercially prepared mixture of $CH_4$ and $H_2$ (2 mole % of $CH_4$) gases was loaded; the Raman spectroscopy results are consistent with other compositions and supports the gas mixture methodology adopted here. Indeed, the overall results show that the experimental Raman and calculated nominal gas compositions show a good proportionality in both fluid $CH_4$-$H_2$ mixture and $((CH_4)_x(H_2S)_{(1-x)})_2H_2$ crystals (Fig. 1) except for the top $CH_4$ concentration, where the gases were likely not uniformly mixed because of insufficient time used (Fig. 1(b)).

**Table 1**. **Diamond anvil cell experiments on $((CH_4)_x(H_2S)_{(1-x)})_2H_2$.**

| Experiment | Partial pressures*) (kpsi) $H_2$; $CH_4$; $H_2$ | Calculated gas composition ($n_{CH4}/n_{total}$) | Raman $CH_4$ composition | Top pressure investigated | Experimental Techniques |
|---|---|---|---|---|---|
| AG004<br>AG005 | 12; 12; 0<br>12; 12; 0 | 0.154(6) | 0.074(8) | 64<br>143 | XRD/Raman/ abs |
| AG007 | 10; 4; 10 | 0.127(6) | 0.094(10) | 20 | Raman |
| AG012 | 10; 2; 12 | 0.075(6) | 0.040(4) | 31 | Raman/abs |
| AG014 | 10; 1; 13 | 0.041(6) | 0.029(4) | 25 | Raman |
| AG015<br>AG016 | 10; 0.5; 13.5<br>10; 0.5; 13.5 | 0.021(5) | 0.020(3) | 34<br>250 | XRD/Raman |
| XZ003 | Mixed before | 0.02 | 0.011(3) | 40 | Raman |

*) Partial pressures of gases, which were loaded in the presented sequence to the total pressure of 24,000 psi.

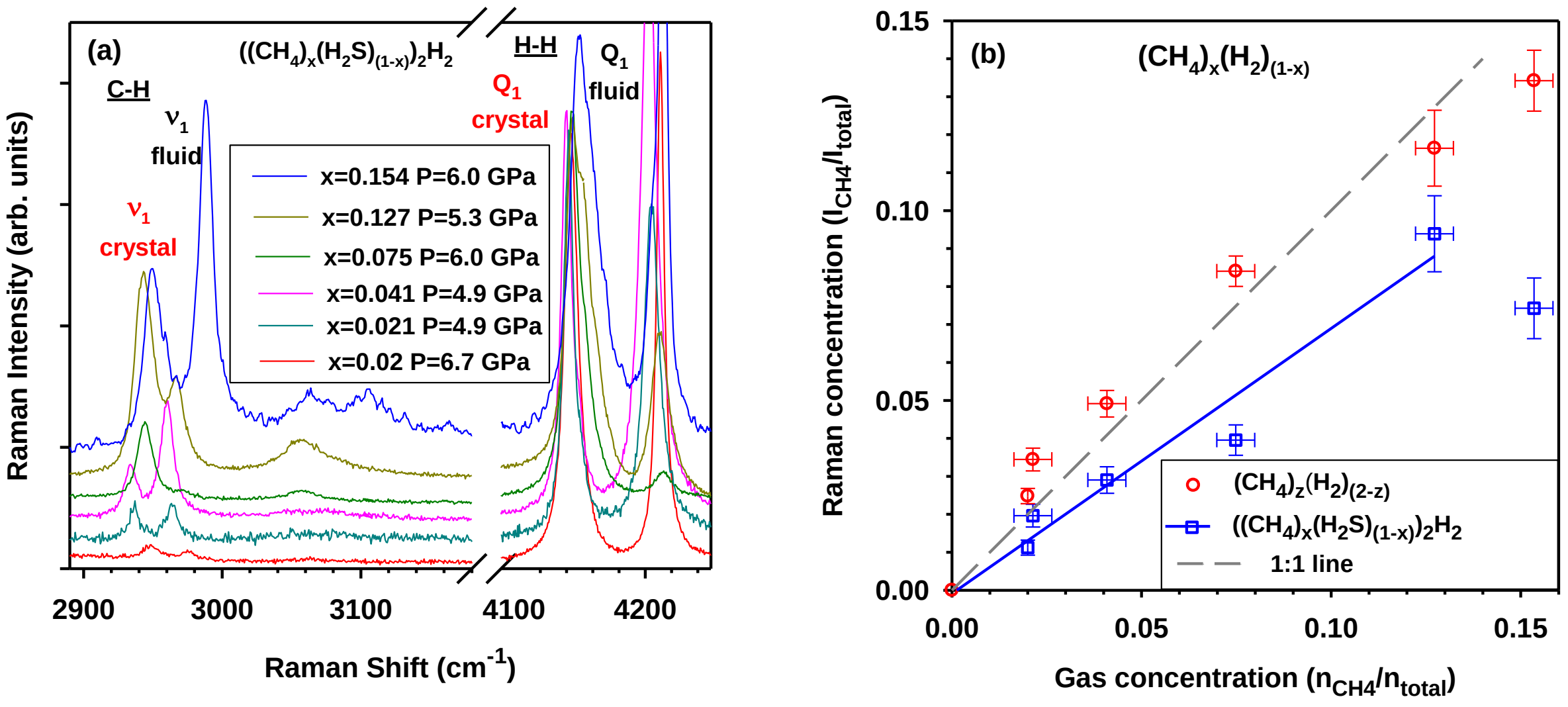

**Figure 1**. Raman spectroscopic determination of $CH_4$ composition in $(CH_4)_x(H_2)_{(1-x)}$ fluids and $((CH_4)_x(H_2S)_{(1-x)})_2H_2$ compounds. (a) Raman spectra of the C-H and H-H stretching modes of $((CH_4)_x(H_2S)_{(1-x)})_2H_2$ compounds (crystal) after synthesis at 4.9-7 GPa. The Raman frequencies of the crystal are red shifted compared to those of surrounding $(CH_4)_x(H_2)_{(1-x)}$ fluid; (b) $CH_4$ compositions in fluid and solid compound determined from the relative Raman peak areas (see Fig. S1 of Supplementary Material [23] for $(CH_4)_x(H_2)_{(1-x)}$ fluids) using the Raman cross sections

determined below 30 MPa [24] vs the gas composition determined from the gas partial pressures (Table 1).

In the experiment AG004 and AG007 (Table 1), we grew the sample at 4 GPa by illuminating S with a tightly focused 457 nm laser with up to 170 mW power as described previously [17]. In other experiments (AG012, 14, 15), after the gas loading, we heated sulfur at 0.5-2.5 GPa using the same laser until it reacted with the gas medium. The reaction continued with time at room temperature and after several hours there was no more elemental sulfur in the cavity, which can be visually observed or detected by Raman spectroscopy. Instead, over time, a vesicle forms in the cavity (Fig. 2), which we identified as $CH_4$ doped $H_2S$ fluid using Raman spectroscopy (Fig. S2 of Supplementary Material [23]), which shows the red shifted C-H and S-H stretch modes and the S-H bend mode. This material crystallizes at 2-3 GPa recognized by its taking a shape with corners (Fig. 2). At above 4.5 GPa this crystal started vanishing and at 4.9 GPa we observed the formation of a new crystal in another close but different location (Fig. 2). Within several tens of minutes, this crystal grew up and that of $CH_4$ doped $H_2S$ completely disappeared. The high-pressure crystal (a second smaller one is also visible) has been identified as $CH_4$ doped $(H_2S)_2H_2$ based on very characteristic Raman spectra of the H-H stretch mode, which shows a peak lower in frequency than that of bulk $H_2$ (Figs. 2 and S2) and XRD experiments as described below. It is important to note that the C-H stretching modes of $((CH_4)_x(H_2S)_{(1-x)})_2H_2$ solid are red shifted with respect to those of $(CH_4)_x(H_2)_{(1-x)}$ fluids, which has been uniquely recognized in this work (see also a previous publication from our group [17]) by Raman measurements with a high spatial resolution, where we were able to collect selectively the spectra of $((CH_4)_x(H_2S)_{(1-x)})_2H_2$ solids almost free of other contributions (Fig. 2 and Fig. S2 of Supplementary Material [23]). Compression above 7 GPa results in formation of $CH_4$ rich and poor regions and subsequent crystallization of $CH_4$-$H_2$ compounds[25, 26] (which can be $H_2S$ doped). These compounds typically form very close to $((CH_4)_x(H_2S)_{(1-x)})_2H_2$ crystals (Fig. 2). The Raman spectra of $CH_4$-$H_2$ compounds have distinct frequencies of the C-H and H-H stretching modes, which are higher in frequencies than those of $((CH_4)_x(H_2S)_{(1-x)})_2H_2$ crystals (Figs. S2-S4 of Supplementary Material [23]). The close proximity at which $((CH_4)_x(H_2S)_{(1-x)})_2H_2$ and $CH_4$-$H_2$ compounds form often results in complex Raman spectra, where the C-H and H-H stretching mode are greatly overlap, making it difficult to identify the unique C-H stretching modes of $((CH_4)_x(H_2S)_{(1-x)})_2H_2$ compound at high pressures unless a careful identification is performed as in this work.

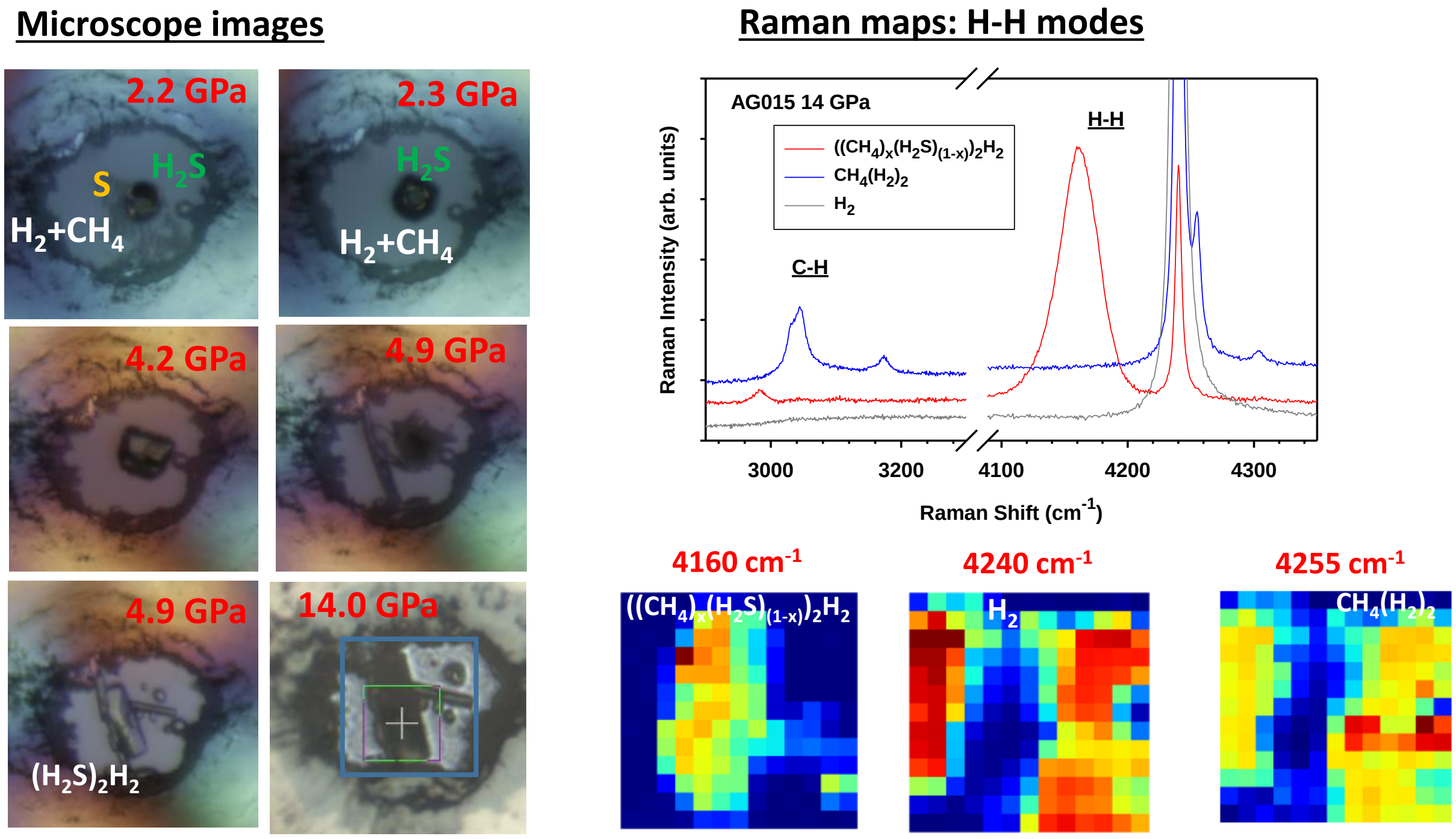

**Figure 2**. Synthesis of $CH_4$ bearing $((CH_4)_x(H_2S)_{(1-x)})_2H_2$ compounds with x=0.02. (a) A series of microscopic images demonstrating a sequence of events related to synthesis of $((CH_4)_x(H_2S)_{(1-x)})_2H_2$ crystalline compound; (b) Raman spectra in the most characteristic sample areas characterizing the typical materials at 14 GPa; (c) Raman maps demonstrating the abundance of materials in the DAC cavity determined at the characteristic frequencies of the H-H stretching modes, corresponding to different species.

For the highest pressure experiment (AG016) we explored a direct synthesis at 132 GPa. In this experiment up to 240 GPa, after gas loading the pressure was 132 GPa and the presence of methane molecules was verified by Raman spectroscopy. The presence of sulfur and gold (pressure calibrant) was verified by using synchrotron powder diffraction. Sulfur sample was laser heated to above 2000 K at the GeoSoilEnviroCARS beamline (Advanced Photon Source, ANL, Chicago) using a double-side near IR fiber laser system combined with synchrotron X-ray diffraction. After sulfur has been reacted with $H_2$ and $CH_4$, a C-S-H compound has been formed similar to that described previously in $CH_4$ free samples [9, 27, 28]. We assumed the presence of $CH_4$ molecules in C-S-H compound based on the stability of this material (albite with a larger $CH_4$ composition) determined previously at pressures up to 143 GPa [17]. This sample (as well the others examined by XRD) has been annealed at each pressure point to increase the crystal quality, reach chemical equilibrium, and to facilitate phase transformations by overcoming the kinetic barriers. Concluding the synthesis section, we state that $((CH_4)_x(H_2S)_{(1-x)})_2H_2$ compounds can be synthesized at a variety of pressures (4-132 GPa) and from different reagents but the final results appear to be independent on the synthesis protocol based on our Raman and XRD investigations (Table 1).

The single-crystal (SC) XRD and Raman spectroscopy experiments have been performed at various pressures up to the highest pressures reached (Table 1). In the AG015 sample, we also released pressure from 34 to 14 GPa attempting to catch the details of superstructure, which has been observed between 20 and 34 GPa. Single-crystal XRD has been measured at each pressure point, while Raman experiments were collected in selected pressure points and in dedicated experiments. Combined XRD and Raman mapping has been used for several pressure points in the AG004, AG005, and AG015 experiment. The details of the data collection, structure determination and refinement are presented in our previous publication [17] and in Supplementary Material [23].

The Raman spectra of the samples synthesized here demonstrate the incorporation of methane molecules for all the concentrations studied (Table 1). The general behavior of the major vibrational modes is similar for materials with different concentration demonstrating the S-H and H-H stretching modes softening with pressure, while the C-H stretching modes harden (see also Refs. [4, 17]). However, there is a difference depending on concentration. While for small concentrations $x \leq 0.04$, Raman spectroscopy demonstrates the presence of the fine structure in the S-H and H-H stretching modes and narrow low-frequency lattice modes above approximately 20 GPa (Fig. 3 and Figs. S3, S4 of Supplementary Material [23]), for crystals with larger $CH_4$ content ($x$=0.074 and 0.094) the Raman spectra remain broad and largely structure less. These observations correlate well with the results of SC XRD measurements (see below), which show the presence of *C*2/*c* structure in $((CH_4)_x(H_2S)_{(1-x)})_2H_2$ compounds with $x \leq 0.04$, representing a distortion of *I*4/*mcm* structure with a 3 times larger unit cell volume. This fact provides a plausible explanation of the complex low-frequency lattice mode spectra in terms of the acoustic bands folding, which results in observations of the phonons at multiple high-symmetry Brillouin zone points, which become the zone center lattice modes of the enlarged unit cell of *C*2/*c* structure. This Brillouin zone folding is also likely responsible for the observations of rich spectra of the S-H and H-H modes (Fig. 3 and Figs. S3, S4 of Supplementary Material [23]). However, this splitting of the vibrational modes persists even in *I*4/*mcm* structure, which restores above 30 GPa, suggesting that orientational order of $H_2$ and $H_2S$ molecules may contribute to the mode splitting in this regime via observation of the crystal field splitting for the molecules in a different crystallographic and hence variable intermolecular force environment.

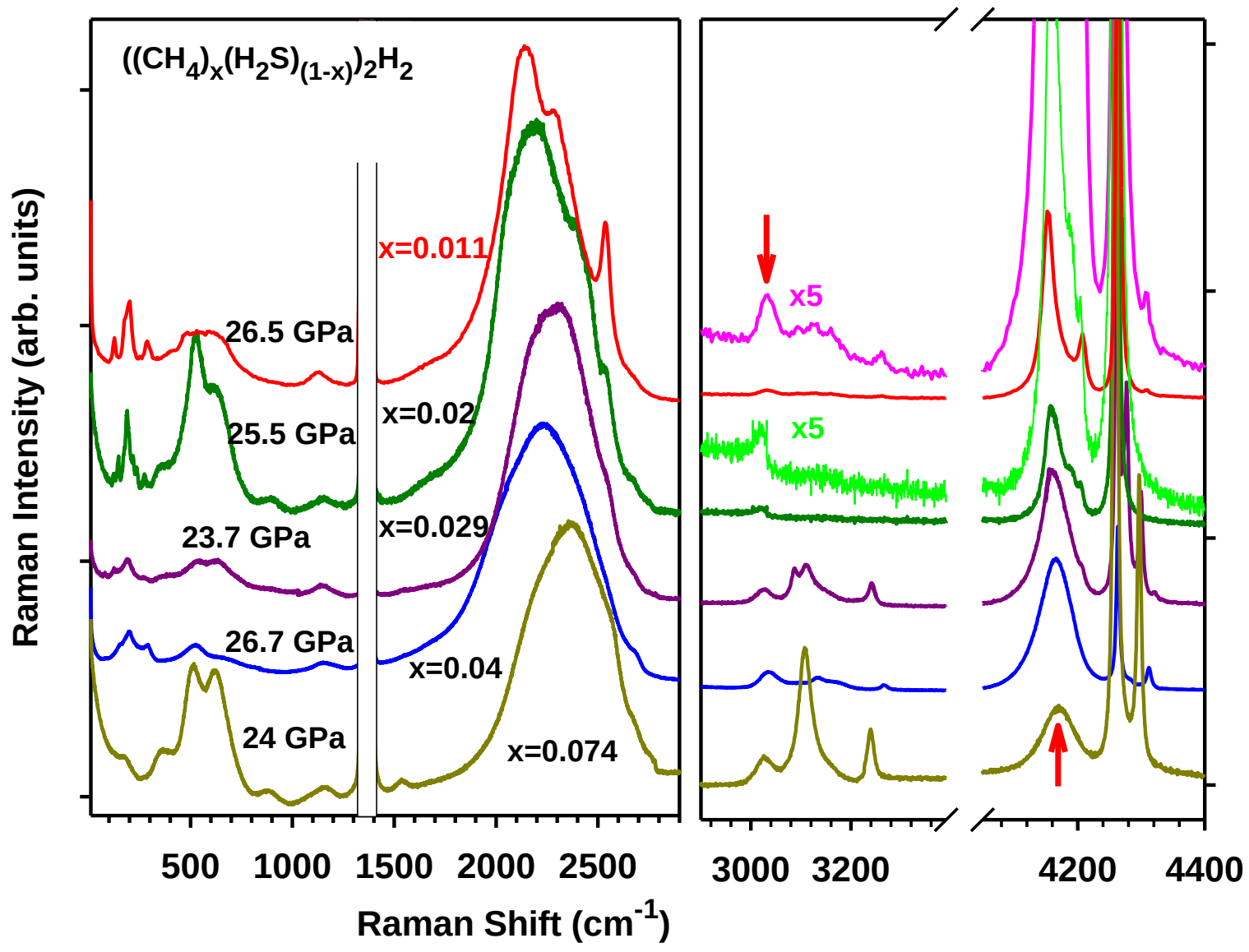

**Figure 3**. Raman spectra of $((CH_4)_x(H_2S)_{(1-x)})_2H_2$ compounds at 24-26.7 GPa for various methane contents x, inferred from the intensity of the C-H stretch mode at low pressures (Fig. 1, Table 1). The laser excitation wavelengths are 532 nm (green and yellow curves, x=0.02 and 0.074), and 488 nm (blue, dark pink, and red curves, x=0.04, 0.029, and 0.011). The red arrows mark the specific red-shifted C-H and H-H stretch modes of $((CH_4)_x(H_2S)_{(1-x)})_2H_2$ compounds, while other modes of the same origin are due to surrounding bulk $H_2$ and $CH_4$-$H_2$ compounds. The first-order Raman signal of diamond anvils is masked by a rectangle.

There is another difference in the high-pressure behavior of $((CH_4)_x(H_2S)_{(1-x)})_2H_2$ with x≤0.04 compared to that of $CH_4$ rich materials. The spectra are very similar at low pressures below 14 GPa, where they contrast only by the difference in the intensity of the C-H stretch modes. However, at higher pressures, the S-H and H-H stretching modes demonstrate a more pronounced softening (Fig. 3 and Fig. S4 of Supplementary Material [23]), while the C-H stretching mode frequencies are the same within the experimental uncertainty. The S-H stretching modes become coupled to the lower frequency S-H bending modes at 35 GPa similar to the case of $H_2O$ [29].

The mode behavior of the S-H bands are well explained by two-stage ordering of $H_2S$ molecules and their polymerization via formation of symmetric hydrogen bonds [7-9, 17]; these phenomena naturally become deteriorated with the increase of $CH_4$ doping (substituting $H_2S$ molecules, see below). The Raman spectra of $((CH_4)_x(H_2S)_{(1-x)})_2H_2$ with x≤0.04 are very similar to those of pure $(H_2S)_2H_2$ [8] and those of the batch B1 (C doped $(H_2S)_2H_2$) from our previous work [17] except for the presence of the C-H stretch modes. Moreover, these spectra are very similar (including also a very similar pressure dependence of the Raman mode frequencies, Fig. 4) to those presented by Snider et al. [4] and assigned to $((CH_4)_x(H_2S)_{(1-x)})_2H_2$ compounds with x close to 0.5 and to those reported more recently with undetermined x by Lamichhane et al. [18] at 4 GPa. This similarity extends to higher pressures above 14 GPa, where our spectra agree well with those reported in Ref. [4] in the behavior of the S-H and H-H stretching modes (Fig. 4) and the presence of a large number of narrow lattice modes (Fig. 3 and Figs. S3-S4 of Supplementary Material [23]). However, there is a difference between our work and Ref. [4] in that we report only two C-H stretching modes in $((CH_4)_x(H_2S)_{(1-x)})_2H_2$ compounds compared to four in Ref. [4]. These four modes of Ref. [4] are very close in frequency to those, which we measured here and assigned to $CH_4$-$H_2$ compounds (Figs. 3, 4, and Figs. S2-S4 of Supplementary Material [23]). On the other hand, Snider *et al.* [4] did not seem to observe the C-H stretching mode, which we unmistakably assign here to the one of $CH_4$ molecule incorporated into $((CH_4)_x(H_2S)_{(1-x)})_2H_2$ compounds (2930 $cm^{-1}$ at 4.5 GPa, see Fig. 4). Please note that the recent work of this group [18], where the spectra with the better signal-to-noise ratio are presented, does reveal such mode as a weak shoulder (Fig. 4). We suggest that yet not reproduced synthesis procedure of Refs. [4, 18] created not fully unidentified hydrocarbon compounds residing in the close proximity to $((CH_4)_x(H_2S)_{(1-x)})_2H_2$ crystals, and their reported C-H spectra are largely due to these compounds. However, given the other similarities in the Raman spectra of $((CH_4)_x(H_2S)_{(1-x)})_2H_2$ compounds between these works and our results, we suggest that the materials synthesized here (see also Ref. [17]) and in works of Refs. [4, 18] are very similar. Based on the doping dependence of the Raman spectra studied here, we can further suggest that doping in the superconducting crystals in Ref. [4] was small, likely comparable to that for the samples with x=0.01-0.04 in our work. This makes our structural study presented here highly relevant for understanding of high-temperature superconductivity.

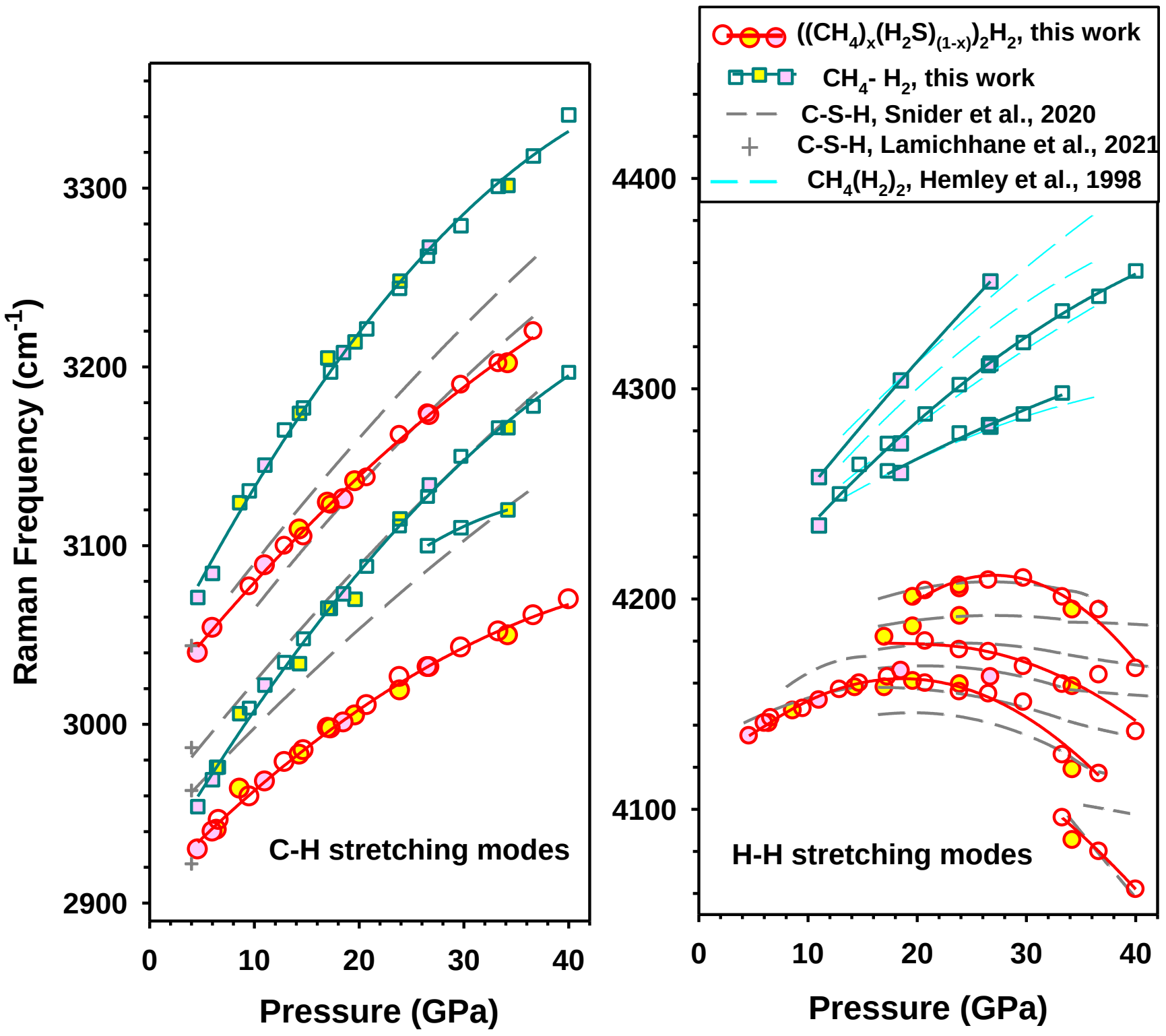

**Figure 4**. Raman frequencies of the C-H and H-H stretching modes of $((CH_4)_x(H_2S)_{(1-x)})_2H_2$ compounds (red symbols) and $CH_4$-$H_2$ compounds (dark cyan symbols). The symbols filled with different colors correspond to different x compositions (Table 1): no filling -0.011, yellow fill-0.02, pink fill-0.04. Dashed gray lines are the data from Snider et al. [4] and gray crosses are from Ref. [18]. Dashed light cyan lines are the results from Ref. [26] in $CH_4$-$H_2$ compounds.

We switch now to structural properties of $((CH_4)_x(H_2S)_{(1-x)})_2H_2$ compounds determined by synchrotron SC XRD. Here we focus on investigations of low doped $((CH_4)_x(H_2S)_{(1-x)})_2H_2$ compound with x=0.02 in experiments AG015 and AG016. Our previously reported results for x=0 and x=0.074 [17] are used for comparison. These experiments demonstrated the same phases (*I*4/*mcm*, *C*2/*c*, *Pnma*, and *Im*-3*m* phase) as have been identified previously [17].

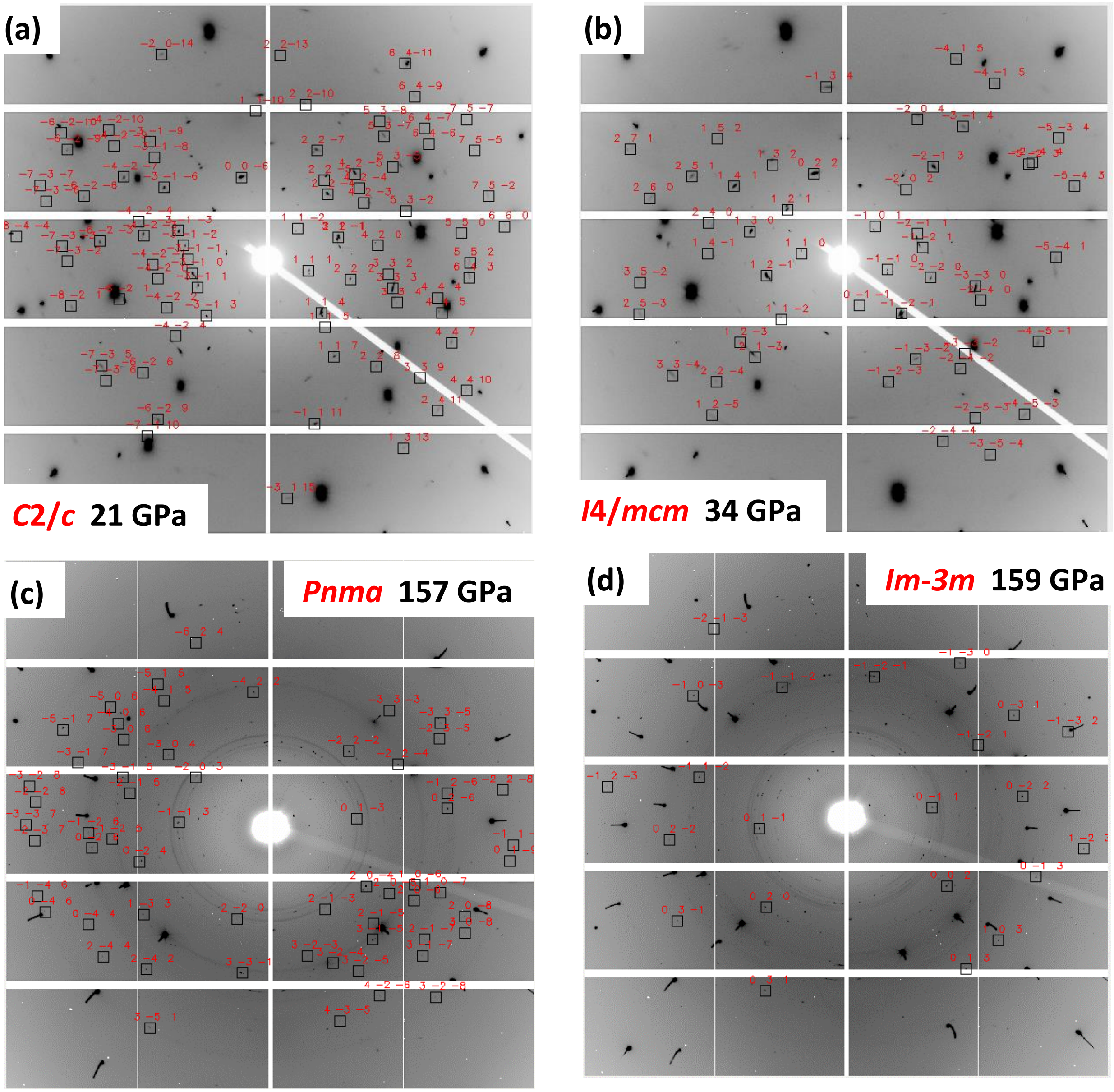

**Figure 5**. Single crystal XRD diffraction at high pressures of the documented phases of $((CH_4)_x(H_2S)_{(1-x)})_2H_2$ (x=0.02). The patterns are taken in a wide scan mode in the range of the ω rotations of ±10° for *C*2/*c* (a), *I*4/*mcm* (b), and *Pnma* (c) phases and ±20° for *Im*-3*m* (d) phase.

To illustrate the data quality we show the representative SC XRD patterns, which were collected as wide scans than spanned in the diffraction angle θ ±10° for *I*4/*mcm*, *C*2/*c*, and *Pnma* phases and

±20° for *Im*-3*m* phase (Fig. 5). The details of the structural refinement of AG015 and AG016 (x=0.02) samples are presented in the Tables S1-S4 of Supplementary Material [23]. Similar results for the materials with x=0 and x=0.074 were presented previously[17]. The experiments demonstrate that single crystals of high quality can be preserved to at least 160 GPa by laser annealing the sample. At higher pressures, the samples breaks into smaller grains, the biggest of them are comparable to the X-ray beam focal spot size. The quality of the data are still sufficient to perform the full structural refinement and extract the structural information as demonstrated in the Fig. 6.

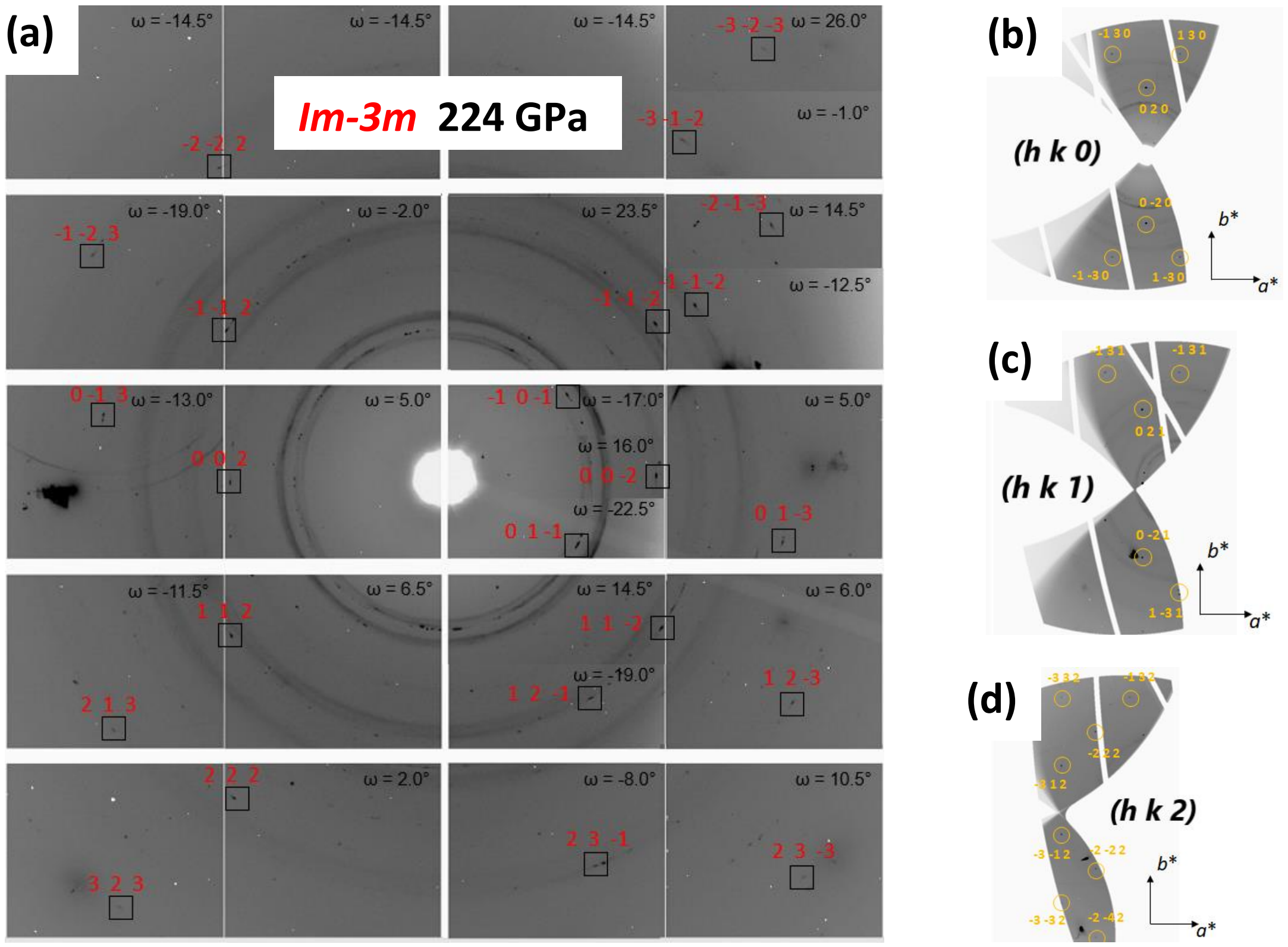

**Figure 6**. Single crystal XRD diffraction of *Im*-3*m* $((CH_4)_x(H_2S)_{(1-x)})_2H_2$ (x=0.02) at 224 GPa. Panel (a) is a mosaic if patterns taken at the different ω rotations angles, demonstrating a variety of *(h k l)* reflections. The panels (b), (c), (d) show *(h k l)* slices of reciprocal space with attributed diffraction spots.

As in our previous work [17], the exact determination of hydrogen positions attached to S atoms was not possible, due to low scattering power of hydrogen compared to sulfur. Also, the substitution of sulfur by carbon does not improve the refinement suggesting that substitutional carbon atoms occupy the same crystallographic positions as sulfurs. The $CH_4$ doped samples investigated here are different from those of our previous work [17] in that the carbon composition is smaller, which results in more ordered structures (e.g., *C*2/*c* at 20-30 GPa in samples with x=0.02), which are essentially indistinguishable from those, which we documented for pure $(H_2S)_2H_2$ compounds in their pressure stability range. The structural stability information is summarized in the Fig. 7, where we show the stable phases of $((CH_4)_x(H_2S)_{(1-x)})_2H_2$ compounds as a function of pressure and $CH_4$ content. Our reported here experiments on $((CH_4)_x(H_2S)_{(1-x)})_2H_2$ compounds with x=0.02 did not reveal any new phase compared to the reported previously for compositions x=0 [9, 10, 17, 30, 31] and x=0.074 [17].

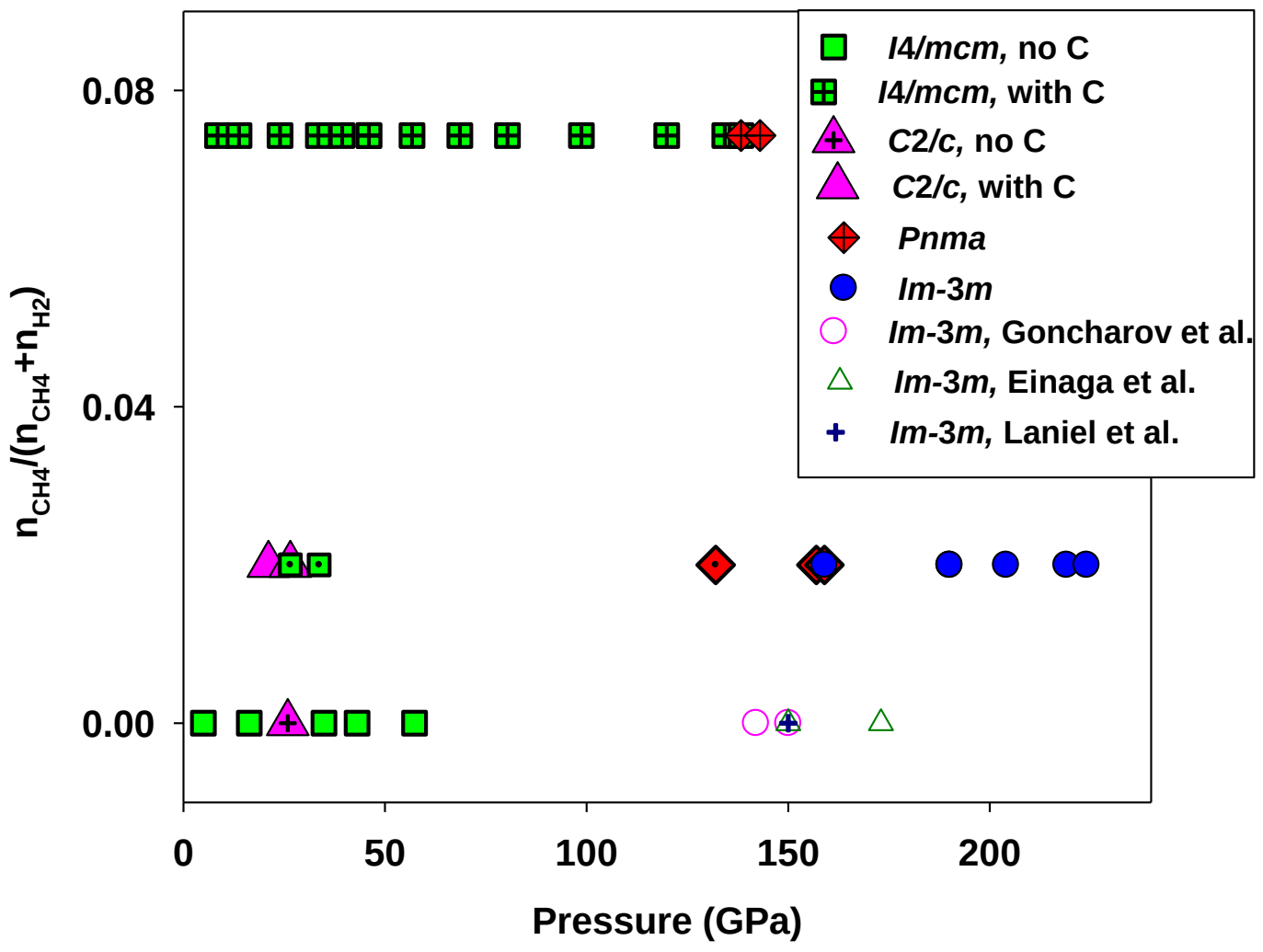

**Figure 7**. The pressure-composition phase diagram of $((CH_4)_x(H_2S)_{(1-x)})_2H_2$ compounds. The structural data of this work are shown with filled symbols, while the stability of *Im*-3*m* $H_3S$ is indicated by open symbols following the results of Refs. [9, 10, 30, 31]. The *R*3*m* $H_3S$ detected at 70-140 GPa is not shown as it is likely metastable [9].

Our experiments combined with the results of experiments of other authors (Refs. [9, 10, 30, 31]) in pure $(H_2S)_2H_2$ show that $CH_4$ doping results in stability of *Pnma* structure above 130 GPa [17] (see also Ref. [18]). This may or may not be related to the reported emergence and raise of $T_c$ in high-temperature superconductor reported in Ref. [4]. Indeed, in our experiments aimed to determine the structure of the stable C-S-H phases we annealed the samples at each pressure points, which could result in promotion of a new physical or even chemical transformation. However, our observations show that *Pnma* structure transforms to *Im*-3*m* at 159 GPa even without temperature annealing demonstrating a phase mixture before laser heating (Figs. 7, 8). This result seems to be in odd with a recently presented XRD results of Ref. [18], which reports the stability of an “orthorhombic phase” and no *Im*-3*m* phase up to 178 GPa. However, one should note that pressure measurements in Ref. [18] were based on the positions of Rhenium and Rhenium hydride diffraction peaks of the gasket, which could produce the large uncertainties (cf. in cavity gold sensor here and in Ref. [17]). Indeed, one can notice that the volume – pressure slope is much smaller in Ref. [18], while the volume values are comparable to those of Ref. [17]. At 174 GPa, which is the top pressure of Lamichhane et al. [18] study, they measured the same unit volume of their “orthorhombic phase” as in our experiment at 159 GPa, where we recorded the transition to *Im*-3*m* phase, thus making both these experiments consistent with regard to the phases observed apart from the apparent difference in the reported pressures.

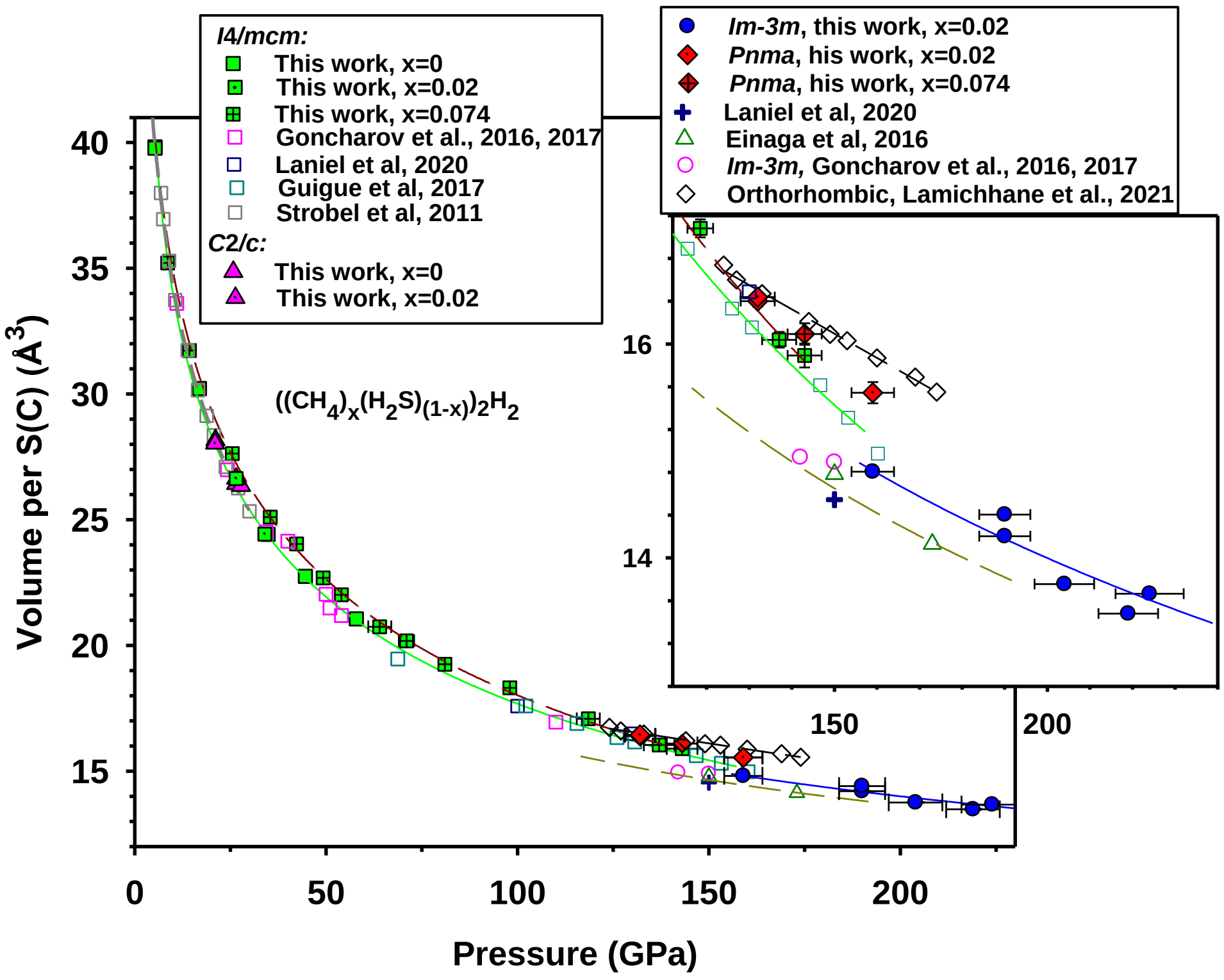

**Figure 8**. The pressure volume equation of state of $((CH_4)_x(H_2S)_{(1-x)})_2H_2$ compounds investigated in this work (see also Ref. [17]) in comparison with other investigations of Refs. [7, 9, 10, 18, 30-32]. The inset shows the details of the high-pressure behavior. A green solid line is an approximation of the *I4/mcm* EOS of $(H_2S)_2H_2$ [7-10, 30, 32] using the Vinet equation of state ($P_0$ =4.8 GPa, $V_0$ = 41.5(1)Å$^3$, $K_0$ = 15.74 GPa, $K_0'$ = 5.11). A brown dashed line is the fit of the *I4/mcm* EOS of $((CH_4)_x(H_2S)_{(1-x)})_2H_2$ compound with x=0.074 of this work using the Vinet equation of state ($P_0$ =4.8 GPa, $V_0$ = 41 (2)Å$^3$, $K_0$ = 22.5(4.0) GPa, $K_0'$ = 4.4(3)). A dashed yellow line is an approximation of the *Im*-3*m* EOS of $H_3S$ [9, 10, 30, 33] using the Birch-Murnagan equation of state ($P_0$ =150 GPa, $V_0$ = 14.65Å$^3$, $K_0$ = 610 GPa, $K_0'$ = 4), while the blue solid line is the Birch-Murnagan equation approximation of the data of this work (the same parameters but $V_0$ = 15.03 Å$^3$ ) for *Im*-3*m* $((CH_4)_x(H_2S)_{(1-x)})_2H_2$ compound with x=0.02. The error bars for determination of the volume are smaller than the symbols size (if not shown).

The pressure volume relations measured in this work for $((CH_4)_x(H_2S)_{(1-x)})_2H_2$ compounds with various x composition provide an insight to the character of $CH_4$ doping and a guidance for XRD characterization of the composition of the high-pressure phases. Doping with $CH_4$ results in swelling of the unit cell depending on the doping concentration and the phase. In the Fig. 9 we present a difference in the unit cell volumes (per the formula unit) between the doped compounds and pristine $(H_2S)_2H_2$ and $H_3S$ for various phases and x concentrations. We used the EOSs of *I4/mcm* $(H_2S)_2H_2$ and *Im*-3*m* EOS of $H_3S$ constructed from this work and the literature data (Fig. 8) as a baseline. Doping of *I4/mcm* $((CH_4)_x(H_2S)_{(1-x)})_2H_2$ with x=0.074 results in a large (up to 1 Å$^3$/f.u.) volume increase at 15-60 GPa, while it drops at higher pressure and becomes comparable to other phases. On the other hand, a smaller doping of x=0.02 leads to a much smaller volume expansion, which is almost within the combined experimental uncertainty. Our Raman spectroscopy results (Fig. 3) show a clear difference in orientational ordering for smaller and larger doping concentrations, which must account for the observed volume difference. The ordered *I4/mcm* structures of $((CH_4)_x(H_2S)_{(1-x)})_2H_2$ with x≤0.04 are more compact compared to the case of the orientatinally disordered structure in the heavier doped materials. The importance of this ordering is further supported by the comparison of volumes of the ordered *C2/c* structures at 20-30 GPa, where the compounds with x=0 and x=0.02 show a small volume difference almost within the experimental scatter. It is also instructive that the compound with x=0.074 composition does not crystallize in the ordered *C2/c* structure (Fig. 7) likely because numerous $CH_4$ molecules disrupt ordering between $H_2S$ molecules.

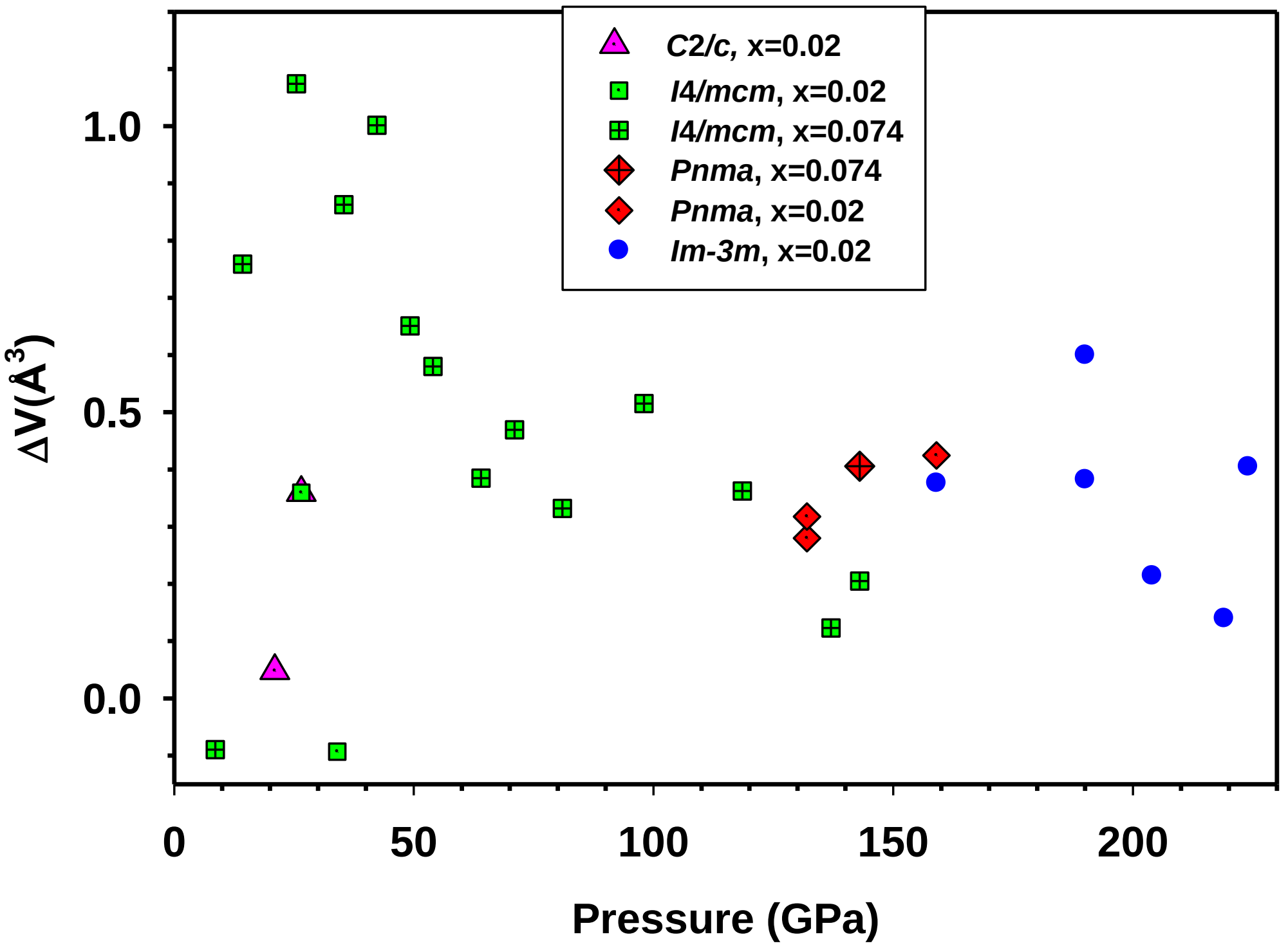

**Figure 9**. The pressure dependencies of the volume difference between the doped and undoped $((CH_4)_x(H_2S)_{(1-x)})_2H_2$ compounds in various phases and with various composition x investigated in this work.

Above 130 GPa, where the extended structures form, we uniquely observe a *Pnma* structure, which is very similar in the positions of sulfur atoms to *I4/mcm*, but is likely largely different concerning the hydrogen positions. Raman spectroscopy shows that the C-H, S-H, and H-H stretching modes vanish above 50 GPa, and only broad low-frequency lattice modes can be seen to higher pressures. On the other hand, the material is likely nonmetal at these conditions [17], suggesting that the drastic changes in Raman spectra are due to changes in the chemical bonding, which results to extended structures similar to observations in $H_2O$ and $H_2$ [29, 34]. The volumes of *Pnma* $((CH_4)_x(H_2S)_{(1-x)})_2H_2$ with x=0.02 and 0.074 are similar, both showing a small expansion compared to the parent *I4/mcm* with the nominally the same composition (see also Ref. [17]). This suggests that this effect is related to the change in structure or in hydrogen content. The positive volume change at the *I4/mcm- Pnma* pressure driven transition seems to be in odd with thermodynamical principles, thus likely signaling an increase in hydrogen content. Given the value of the volume expansion [35], we propose that two additional hydrogen atoms could be accommodated in *Pnma* $((CH_4)_x(H_2S)_{(1-x)})_2H_2$. For the doping values in the range of this work, we expect the composition of our *Pnma* compound would be $S_{15}CH_{68}$. Other much less likely scenarios that would explain the lattice expansion of *Pnma* $((CH_4)_x(H_2S)_{(1-x)})_2H_2$ are a strong structural disorder or a change in S-C stoichiometry. The electronic properties of *Pnma* $((CH_4)_x(H_2S)_{(1-x)})_2H_2$ would drastically depend on hydrogen positions and stoichiometry. Due to a large number of the atoms, the structure is difficult to find theoretically using the state of the art structure prediction tools (e.g. USPEX [36]) thus presenting a challenging task for the future investigations.

As in the case of the parent carbon free material, $((CH_4)_x(H_2S)_{(1-x)})_2H_2$ transforms under pressure to rather simple extended covalent *Im-3m* structure (Figs. 6-8). The *I4/mcm* (*Pnma*) - *Im-3m* transition is related to a substantial atomic reconstruction. Thus, compressing *I4/mcm* at low to room temperature causes the transition shift to higher pressure (up 160 GPa) due to kinetic effects and thus leads to an extended metastability range [8, 32]. By applying laser heating, the transition to the stable *Im-3m* phase can be completed at lower pressures, 140-150 GPa [9, 17]. The transition is associated with a volume contraction, reported to be 8% in Ref. [30] and determined to be about 5% at 150 GPa here based on an accurate extrapolation of the EOS of *I4/mcm*. Our experiments on the doped $((CH_4)_x(H_2S)_{(1-x)})_2H_2$ system with x=0.02 show an increased by approximately 0.35 $Å^3$/f.u. volume compared to the parent $H_3S$ system. This is approximately one order of magnitude larger than one would expect based on an increase by 2% of the number of hydrogen atoms due to the substitution of $H_3S$ by $CH_4$ assuming that one extra hydrogen atom occupies about 2 $Å^3$ [35]. This result is in odd with the theoretical predictions for the ordered $CSH_7$ compounds with a much larger $CH_4$ composition [12, 13]. This suggests that the volume expansion of our slightly doped $((CH_4)_x(H_2S)_{(1-x)})_2H_2$ compounds is due to a lattice disorder. This is in a qualitative agreement with the recent calculations [14, 16], which show an increase in enthalpy of lightly carbon doped $H_3S$ compounds.

Finally, we address the relevance of the presented results to room-temperature superconductivity reported in Ref. [4]. Based on the Raman and XRD results, the synthesized here materials have the structure and composition of carbonaceous sulfur hydride reported in Refs. [4, 18]. Indeed, our experiments are consistent with their XRD measurements concerning the observed phase symmetry and transformation sequence (tetragonal-to-orthorhombic) at 5 -160 GPa. Our Raman results agree in the sequence of the appearance of orientation order above 15 GPa, emergence of extremely rich Raman spectra at 20-30 GPa suggesting an additional superstructure formation, and finally in the high-frequency spectra vanishing due to the formation of an extended structure. We show here that these transformation sequence is sensitive to the $CH_4$ content and is different for compounds with more than 4% doping suggesting that room-temperature superconductivity is a property of the lightly doped compounds. Given a small amount of $CH_4$ in carbonaceous sulfur hydrides, observations of the characteristic C-H Raman modes is intricate. Here we find that the intrinsic C-H modes of $CH_4$ incorporated into $((CH_4)_x(H_2S)_{(1-x)})_2H_2$ compounds are shifted down in frequency compared to bulk $CH_4$ and $CH_4$-$H_2$ compounds. For small x, there are only two local modes corresponding to $\nu_1$ and $\nu_2$ fundamentals of $CH_4$ molecule. Here, we clearly identified these modes and assigned the other C-H stretch modes in the same spectral range to $CH_4$-$H_2$ compounds [25, 26]. Snider et al. [4] observed slightly different C-H stretching mode in their Raman spectra, which do not show the $\nu_1$ mode of incorporated into the C-S-H lattice methane, which contradicts to the $CH_4$-$H_2S$ mixing hypothesis as the result of their photochemistry syntheses. However, recently Lamichhane et al. demonstrated the Raman spectrum at 4 GPa with an improved signal-to-noise ratio, which does show the presence of the $\nu_1$ mode of incorporated into the C-S-H lattice methane demonstrating that their C-S-H crystals do have such molecules similar to the material synthesized here. Therefore, we conclude that the materials investigated here and in Refs. [4, 18] are very similar even though the synthesis protocol is different. Thus, the structural data presented here are relevant for high-temperature superconductivity. In particular, the discovered here *Pnma* phase (see also Ref. [17]), which is stable above 130 GPa and can be only observed for $CH_4$ doped materials (Fig. 7), could be responsible for the initial raise of superconductivity in Ref. [4]. Theoretical calculations, which reproduce this complex structure including hydrogen atom positions, would be highly desirable to test this hypothesis. The highest $T_c$ in the C-S-H systems are likely associated with *Im*-3*m* methane doped $H_3S$, where an increase of $T_c$ is predicted theoretically [14, 15] (cf. Ref. [16]). Our experiments show the stability of such doped materials above 160 GPa; the doping is confirmed by the lattice expansion compared to a parent pure $H_3S$ material (Figs. 8-9). The difference in the transition pressure ($T_c$ upturn was reported at 220 GPa in Ref. [4]) could be due to the difference in the experimental procedure. In our experiments, we applied laser heating which facilitates the phase transformation [9, 27, 30].

In conclusion, we demonstrate here a new straightforward synthesis pathway to create $((CH_4)_x(H_2S)_{(1-x)})_2H_2$ compounds with a variable and controllable composition. We find that room-temperature superconductors emerge in lightly doped materials ($x \leq 0.04$). Our experiments demonstrate the phase sequence *I4/mcm* – *C*2/c – *Pnma* - *Im*-3*m*, which depends on the $CH_4$ composition. We relate the two-stage superconductivity upraise to the *Pnma* - *Im*-3*m* phase change. We propose that laser heating, which facilitate this phase transition would increase $T_c$ and likely reach almost room temperature below 200 GPa. Future experiments will determine the optimal $CH_4$ content, which can be tuned as described in this work.

We thank Timothy Strobel for ball milling of carbon for one of our synthesis attempts. We thank Maddury Somayazulu for information about the Raman intensities in the $CH_4$-$H_2$ system. We thank Dmitrii Semenok, Ivan Kruglov and Artem Oganov for useful discussions about the correspondence of experimental and theoretical structures and properties. Parts of this research were carried out at the GeoSoilEnviroCARS (The University of Chicago, Sector 13), Advanced Photon Source (Argonne National Laboratory). GeoSoilEnviroCARS is supported by the National Science Foundation - Earth Sciences (EAR - 1634415) and Department of Energy GeoSciences (DE-FG02-94ER14466). The Advanced Photon Source is a U.S. Department of Energy (DOE) Office of Science User Facility operated for the DOE Office of Science by Argonne National Laboratory under Contract No. DE-AC02-06CH11357. We acknowledge support by the Army Research Office accomplished under the Cooperative Agreement Number W911NF-19-2-0172 and Carnegie Institution of Washington. Part of this research in China was supported by the National Natural Science Foundation of China (Grant Nos. 51672279, 51727806, 11874361, and 11774354), the CASHIPS Director's Fund (Grant Nos. YZJJ201705 and YZJJ2020QN22), and Science Challenge Projects (Nos. TZ2016001 and TZ2016001).

**Data and materials availability:** The data that support the plots within this paper and other findings of this study are available from the corresponding authors upon reasonable request.

1. A. P. Drozdov, M. I. Eremets, I. A. Troyan, V. Ksenofontov and S. I. Shylin, Nature **525** (7567), 73-76 (2015).
2. M. Somayazulu, M. Ahart, A. K. Mishra, Z. M. Geballe, M. Baldini, Y. Meng, V. V. Struzhkin and R. J. Hemley, Physical Review Letters **122** (2), 027001 (2019).
3. A. P. Drozdov, P. P. Kong, V. S. Minkov, S. P. Besedin, M. A. Kuzovnikov, S. Mozaffari, L. Balicas, F. F. Balakirev, D. E. Graf, V. B. Prakapenka, E. Greenberg, D. A. Knyazev, M. Tkacz and M. I. Eremets, Nature **569** (7757), 528-531 (2019).
4. E. Snider, N. Dasenbrock-Gammon, R. McBride, M. Debessai, H. Vindana, K. Vencatasamy, K. V. Lawler, A. Salamat and R. P. Dias, Nature **586** (7829), 373-377 (2020).
5. E. Snider, N. Dasenbrock-Gammon, R. McBride, X. Wang, N. Meyers, K. V. Lawler, E. Zurek, A. Salamat and R. P. Dias, Physical Review Letters **126** (11), 117003 (2021).
6. P. P. Kong, V. S. Minkov, M.A.Kuzovnikov, S.P.Besedin, A. P. Drozdov, S. Mozaffari, L. Balicas, F.F. Balakirev, V. B. Prakapenka, E. Greenberg, D. A. Knyazev and M. I. Eremets, ArXiv190910482 Cond-Mat (2019).
7. T. A. Strobel, P. Ganesh, M. Somayazulu, P. R. C. Kent and R. J. Hemley, Physical Review Letters **107** (25), 255503 (2011).
8. E. J. Pace, X.-D. Liu, P. Dalladay-Simpson, J. Binns, M. Peña-Alvarez, J. P. Attfield, R. T. Howie and E. Gregoryanz, Physical Review B **101** (17), 174511 (2020).
9. A. F. Goncharov, S. S. Lobanov, V. B. Prakapenka and E. Greenberg, Physical Review B **95** (14), 140101 (2017).
10. D. Laniel, B. Winkler, E. Bykova, T. Fedotenko, S. Chariton, V. Milman, M. Bykov, V. Prakapenka, L. Dubrovinsky and N. Dubrovinskaia, Physical Review B **102** (13), 134109 (2020).
11. D. Duan, Y. Liu, F. Tian, D. Li, X. Huang, Z. Zhao, H. Yu, B. Liu, W. Tian and T. Cui, Sci. Rep. **4**, 6968 (2014).
12. W. Cui, T. Bi, J. Shi, Y. Li, H. Liu, E. Zurek and R. J. Hemley, Physical Review B **101** (13), 134504 (2020).
13. Y. Sun, Y. Tian, B. Jiang, X. Li, H. Li, T. Iitaka, X. Zhong and Y. Xie, Physical Review B **101** (17), 174102 (2020).
14. S. X. Hu, R. Paul, V. V. Karasiev and R. P. Dias, arXiv:2012.10259v1 [cond-mat.supr-con] (2021).
15. Y. Ge, F. Zhang, R. P. Dias, R. J. Hemley and Y. Yao, Materials Today Physics **15**, 100330 (2020).
16. T. Wang, Motoaki Hirayama, Takuya Nomoto, Takashi Koretsune, Ryotaro Arita and J. A. Flores-Livas, rXiv:2104.03710v2 [cond-mat.supr-con] (2021).

17. E. Bykova, M. Bykov, S. Chariton, V. B. Prakapenka, K. Glazyrin, A. Aslandukov, A. Aslandukova, G. Criniti, A. Kurnosov and A. F. Goncharov, Phys. Rev. B **103**, L140105 (2021).
18. A. Lamichhane and M. A. Ravhi Kumar, Nilesh P. Salke, Nathan Dasenbrock-Gammon, Elliot Snider, Yue Meng, Barbara Lavina, Stella Chariton, Vitali B. Prakapenka, Maddury Somayazulu, Ranga P. Dias, Russell J. Hemley, arXiv:2105.06352v2 [cond-mat.supr-con] (2021).
19. S. Duwal and C.-S. Yoo, The Journal of Physical Chemistry C **121** (23), 12863-12870 (2017).
20. I. Kantor, V. Prakapenka, A. Kantor, P. Dera, A. Kurnosov, S. Sinogeikin, N. Dubrovinskaia and L. Dubrovinsky, Review of Scientific Instruments **83** (12), 125102 (2012).
21. R. Boehler, Review of Scientific Instruments **77** (11), 115103 (2006).
22. A. Dewaele, P. Loubeyre, F. Occelli, O. Marie and M. Mezouar, Nature Communications **9** (1), 2913 (2018).
23. See Supplementary Material for Materials and Methods, Tables S1-S3, and Bibliography with 18 References.
24. J. Fang, I.-M. Chou and Y. Chen, Journal of Raman Spectroscopy **49** (4), 710-720 (2018).
25. M. S. Somayazulu, L. W. Finger, R. J. Hemley and H. K. Mao, Science **271** (5254), 1400-1402 (1996).
26. R. J. Hemley, M. S. Somayazulu, A. F. Goncharov and H. K. Mao, Asian J. Phys. **7**, 319-322 (1998).
27. V. S. Minkov, V. B. Prakapenka, E. Greenberg and M. I. Eremets, Angewandte Chemie International Edition **59** (43), 18970-18974 (2020).
28. S. Mozaffari, D. Sun, V. S. Minkov, A. P. Drozdov, D. Knyazev, J. B. Betts, M. Einaga, K. Shimizu, M. I. Eremets, L. Balicas and F. F. Balakirev, Nature Communications **10** (1), 2522 (2019).
29. A. F. Goncharov, V. V. Struzhkin, H.-k. Mao and R. J. Hemley, Physical Review Letters **83** (10), 1998-2001 (1999).
30. A. F. Goncharov, S. S. Lobanov, I. Kruglov, X.-M. Zhao, X.-J. Chen, A. R. Oganov, Z. Konôpková and V. B. Prakapenka, Physical Review B **93** (17), 174105 (2016).
31. M. Einaga, M. Sakata, T. Ishikawa, K. Shimizu, M. I. Eremets, A. P. Drozdov, I. A. Troyan, N. Hirao and Y. Ohishi, Nature Physics **12** (9), 835-838 (2016).
32. B. Guigue, A. Marizy and P. Loubeyre, Physical Review B **95** (2), 020104 (2017).
33. M. Einaga, M. Sakata, T. Ishikawa, K. Shimizu, M. I. Eremets, A. P. Drozdov, I. A. Troyan, N. Hirao and Y. Ohishi, Nat Phys **12** (9), 835-838 (2016).
34. R. T. Howie, C. L. Guillaume, T. Scheler, A. F. Goncharov and E. Gregoryanz, Physical Review Letters **108** (12), 125501 (2012).
35. V. Struzhkin, B. Li, C. Ji, X.-J. Chen, V. Prakapenka, E. Greenberg, I. Troyan, A. Gavriliuk and H.-k. Mao, Matter and Radiation at Extremes **5** (2), 028201 (2020).
36. A. R. Oganov, Y. M. Ma, A. O. Lyakhov, M. Valle and C. Gatti, Theoretical and Computational Methods in Mineral Physics: Geophysical Applications **71**, 271-298 (2010).

**Supplementary Material**

# Synthesis and structure of carbon doped $H_3S$ compounds at high pressure

Alexander F. Goncharov[1], Elena Bykova[1], Maxim Bykov[1,2], Xiao Zhang[3], Yu Wang[3], Stella Chariton[4], Vitali B. Prakapenka[4],

[1] *Earth and Planets Laboratory, Carnegie Institution of Washington, 5251 Broad Branch Road NW, Washington, DC 20015, USA*

[2] *Department of Mathematics, Howard University, Washington, DC 20059, USA*

[3] *Key Laboratory of Materials Physics, Institute of Solid State Physics, HFIPS, Chinese Academy of Sciences, Hefei 230031, Anhui, People's Republic of China*

[4] *Center for Advanced Radiations Sources, University of Chicago, Chicago, Illinois 60637, USA*

## Experimental methods

### 1. X-ray diffraction (XRD) data collection

The XRD measurements were conducted at the 13-IDD (Pilatus CdTe 1M detector, λ = 0.29520 Å, KB-mirror focusing) and 16-IDB beamline (Pilatus Si 1M detector, λ = 0.34453 Å, KB-mirror focusing) at the Advanced Photon Source (APS), Chicago, USA. Sample-to-detector distance, coordinates of the beam center, tilt angle and tilt plane rotation angle of the detector images were calibrated using $LaB_6$ (13-IDD experiments) or $CeO_2$ (13-IDB experiments) powders. The single-crystal XRD images were recorded while rotating the sample about a single ω-axis from -30 to +30° in small steps of 0.5°.

Laser-heating experiments were carried out on a state-of-the art stationary double-side laser-heating setup installed at IDD-13 beamline[1]. The temperature was measured by the standard spectroradiometry method. Here, we report the room temperature XRD data.

### 2. XRD data processing

DIOPTAS software[2] was used for preliminary analysis and for integration of the 2-dimentional images to 1-dimentional diffraction patterns.

Processing of single-crystal XRD data (the unit cell determination and integration of the reflection intensities) was performed using CrysAlisPro software[3]. Empirical absorption correction was applied using spherical harmonics, implemented in the SCALE3 ABSPACK scaling algorithm, which is included in the CrysAlisPro software. A single crystal of an orthoenstatite $((Mg_{1.93},Fe_{0.06})(Si_{1.93},Al_{0.06})O_6$, *Pbca*, *a* = 18.2391(3), *b* = 8.8117(2), *c* = 5.18320(10) Å), was used to calibrate instrument model of CrysAlisPro software (sample-to-detector distance, the detector's origin, offsets of the goniometer angles and rotation of the X-ray beam and the detector around the instrument axis). The calibration crystal was measured in a diamond anvil cell without pressure medium in similar experimental conditions as the studied samples (narrow slicing mode, rotation about ω-axis from -32 to +32° with 0.5°step size).

### 3. Structure solution and refinement.

We have used crystal structures of *C*2/*c*, *Pnma*, and *Im*-3*m* phases described earlier in Refs.[4,5] as starting models for the refinement. The refinement procedures for *C*2/*c* and *Pnma* phases were similar to those provided in our previous work [4]. No definite conclusions could be made about positions of hydrogens in for *C*2/*c* and *Pnma* structures due to low scattering power of hydrogens compared to sulfur. In *Im*-3*m* $H_3S$, the positions of both hydrogen and sulfur atoms are fixed and no coordinates have to be refined: S1 is located on Wyckoff position 2*a* (0, 0, 0) and H1 is on 6*b* (0, 0, 0.5). Only the thermal parameter of sulfur was included into refinement, while thermal parameter of hydrogen was fixed to be 1.5 of this value. Carbon atoms likely remain on sulfur positions for all structures. Occupancy of C would be coupled with the scale factor, therefore carbon was not introduced into the refinement. The crystal structures were refined against $F^2$ on all data by full-matrix least squares with the SHELXL software[6] implemented in Olex2 software package[7].

Unit cell parameters, structural data and details of the structure refinements are given in Supplementary Tables S1-S4.

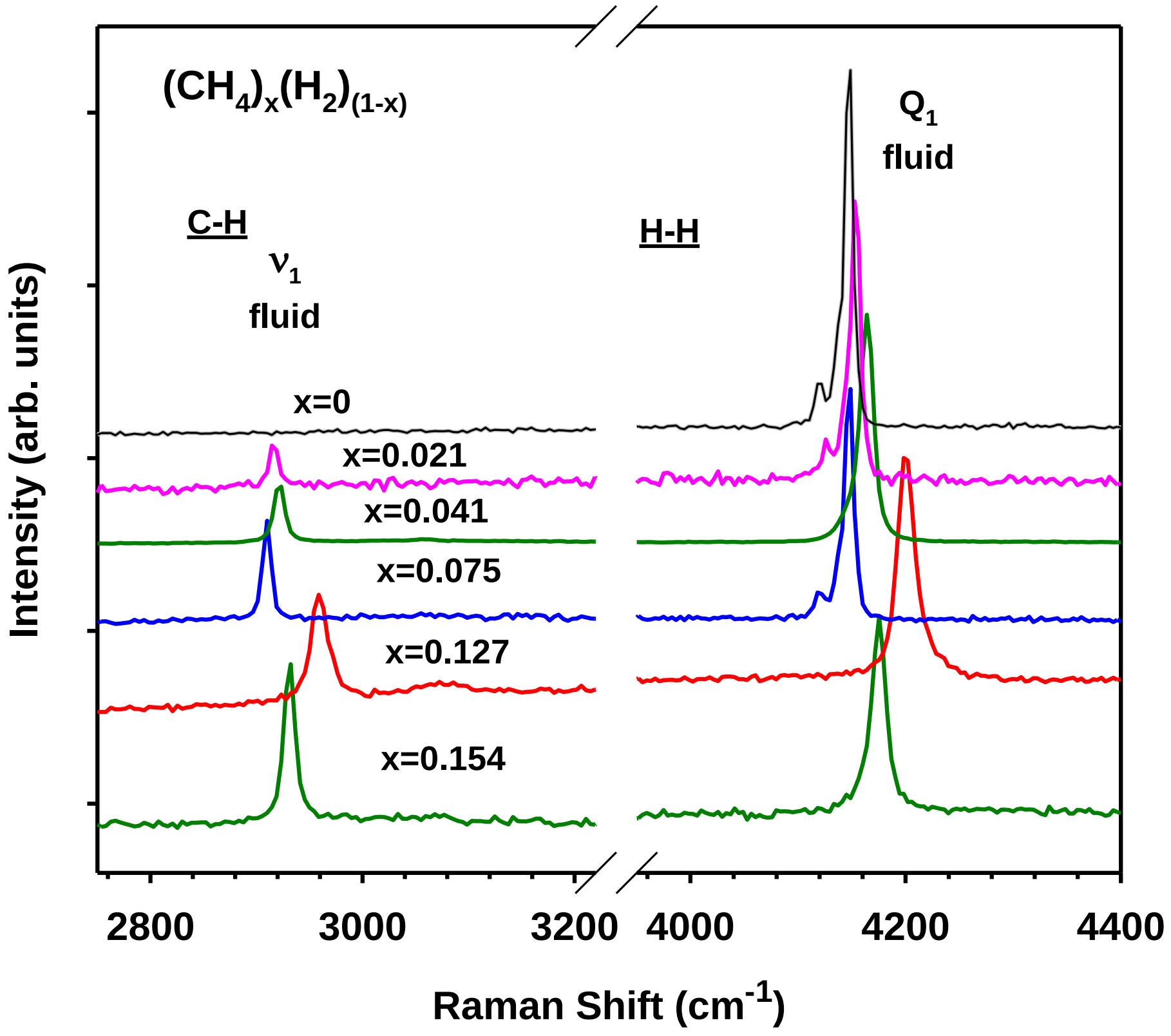

**Figure S1**. Raman spectra of the C-H and H-H stretch modes of $(CH_4)_x(H_2)_{(2-x)}$ fluids after gas loading in the DAC, which served for determination of the $CH_4$ content (Fig. 1(b)).

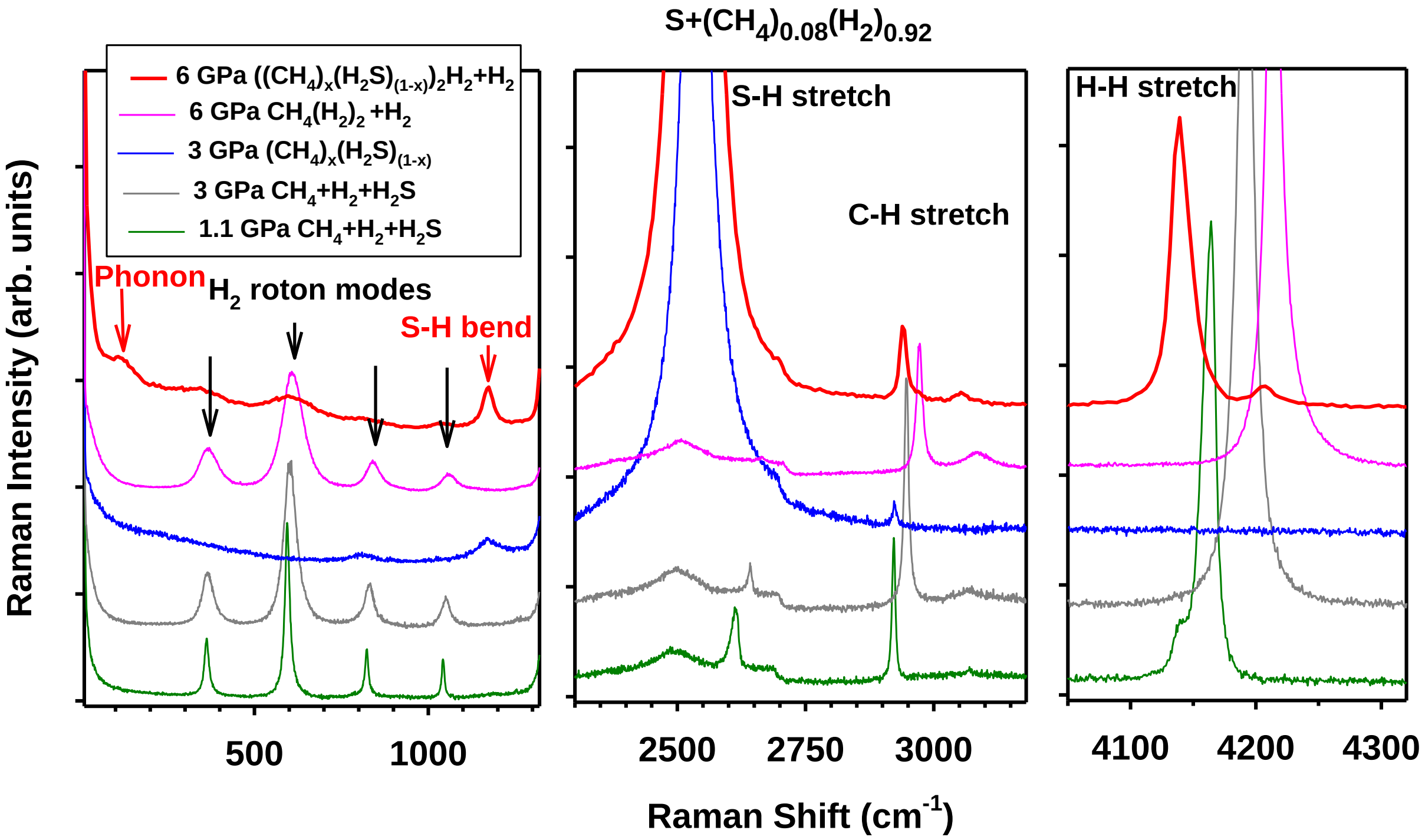

**Figure S2**. Raman spectra of the $(CH_4)_x(H_2)_{(2-x)}$ fluids reacted with S at 1.1 GPa. $H_2S$ liquid forms manifested by the emergence of the S-H stretch modes at 2620 cm$^{-1}$. On the pressure increase a $CH_4$ doped $H_2S$ crystal forms at 3 GPa; this is signified by the red shifted C-H and S-H stretch modes, S-H bend mode, and weak lattice modes. The rest of the sample remains in a mixed fluid state including $H_2$, $H_2S$ and $CH_4$ species. At 6 GPa a $((CH_4)_x(H_2S)_{(1-x)})_2H_2$ crystal forms manifested by the appearance of the red shifted $H_2$ vibron mode and a low-frequency phonon. The C-H and S-H stretch modes remain red shifted compared to the frequencies in the $H_2S$ doped $(CH_4)_x(H_2)_{(2-x)}$ fluids.

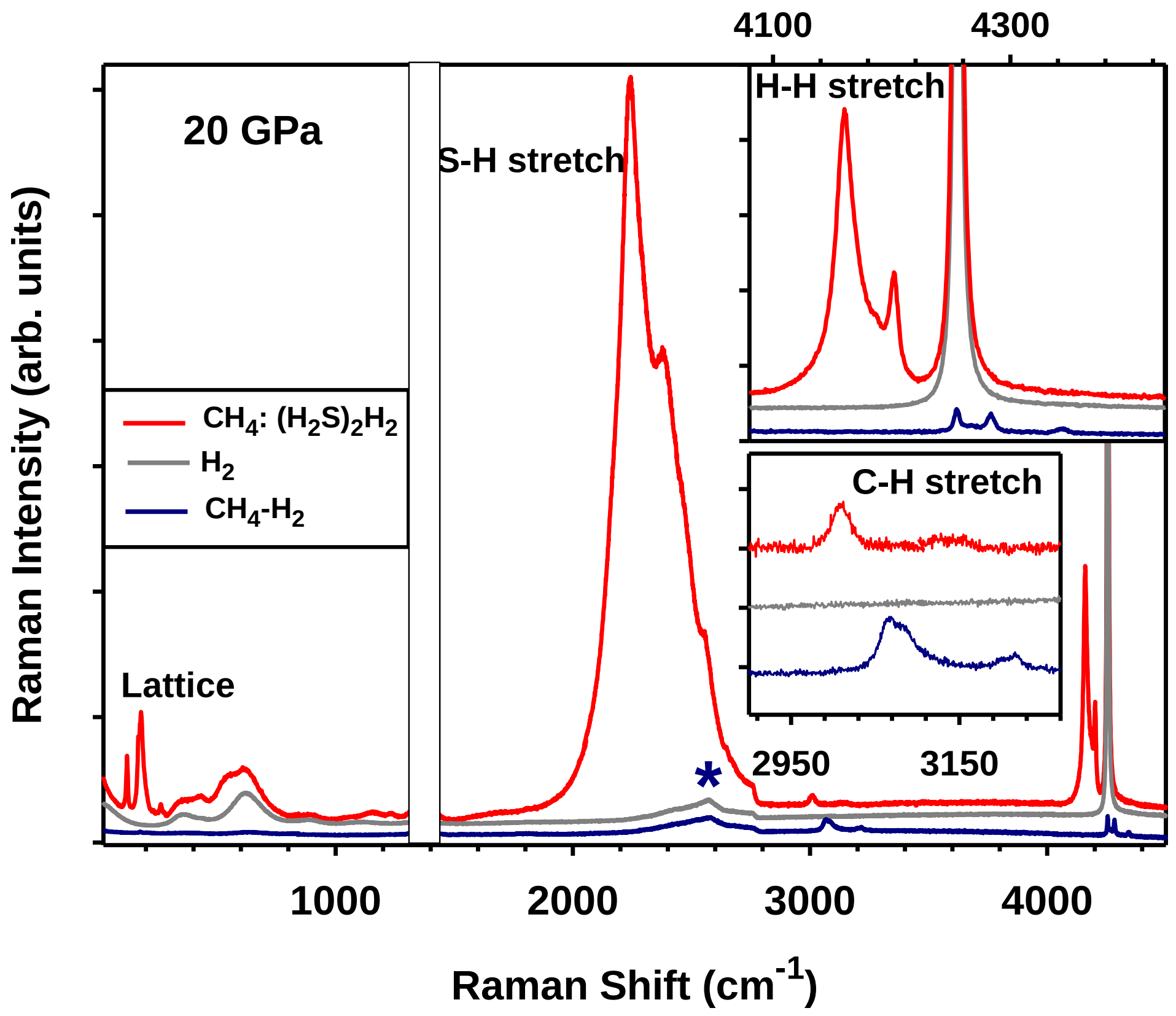

**Supplemental Figure 3**. Raman spectra of the reaction products of S and $(CH_4)_x(H_2)_{(1-x)}$ fluid mixture (x=0.02) after laser heating at 2 GPa and pressure increase to 20 GPa (Fig. S3). The first- and second-order Raman signal of diamond anvils are masked by a rectangle, and marked by an asterisk, respectively. The insets show a zoomed in spectrum of the C-H stretch and H-H stretch modes, which are distinct for a $((CH_4)_x(H_2S)_{(1-x)})_2H_2$ crystal and $CH_4$-$H_2$ inclusion compound.

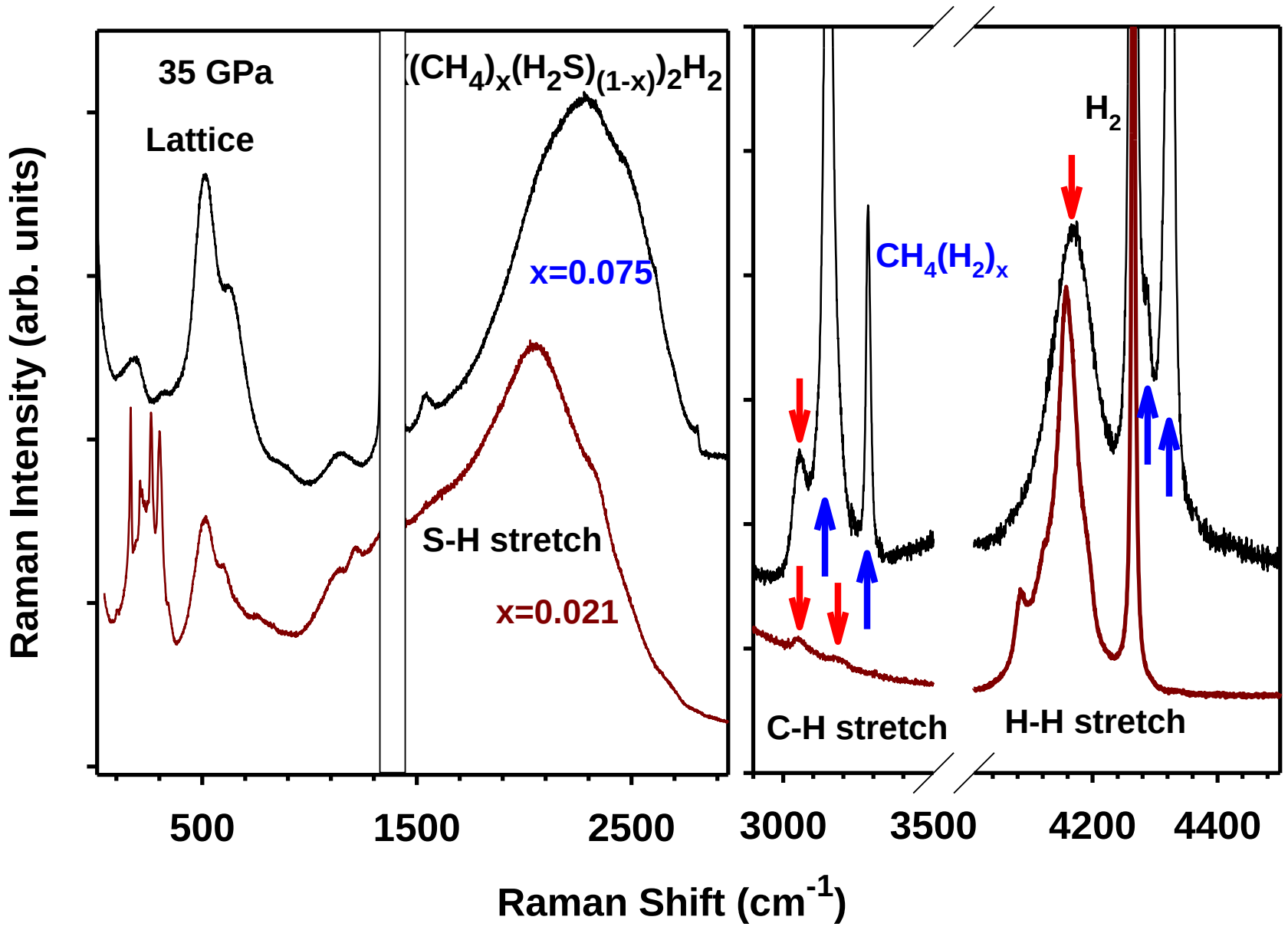

**Supplemental Figure 4**. Raman spectra of $((CH_4)_x(H_2S)_{(1-x)})_2H_2$ compounds at 35 GPa for two methane contents x (Table 1). The laser excitation wavelength is 532 nm. The red arrows mark the specific red-shifted C-H and H-H stretch modes of $((CH_4)_x(H_2S)_{(1-x)})_2H_2$ compounds, while the blue arrows and $H_2$ stand to mark the other modes of the same origin due to surrounding $CH_4$-$H_2$ compounds and bulk $H_2$, respectively The first-order Raman signal of diamond anvils is masked by a rectangle.

## Table S1. Details of crystal structure refinement for *C*2/*c* phase at 21 GPa

| **Sample name** | **AG003_p02_s2** |
|---|---|
| **Pressure, GPa** | 21.1(5) |
| ***a*** **(Å)** | 8.399(6) |
| ***b*** **(Å)** | 6.474(7) |
| ***c*** **(Å)** | 12.503(5) |
| **β (°)** | 98.34(4) |
| ***V*** **(Å$^3$)** | 672.6(9) |
| ***Z*** | 12 |
| **$2\Theta_{min}$ for data collection (°)** | 1.933 |
| **$2\Theta_{max}$ for data collection (°)** | 17.111 |
| **Completeness to *d* = 0.8 Å** | 0.536 |
| **Reflections collected** | 737 |
| **Independent reflections** | 507 |
| **Independent reflections [$I > 2\sigma(I)$]** | 166 |
| **Refined parameters** | 28 |
| **$R_{int}(F^2)$** | 0.0135 |
| **$R(\sigma)$** | 0.0272 |
| **$R_1$ [$I > 2\sigma(I)$]** | 0.0931 |
| **$wR_2$ [$I > 2\sigma(I)$]** | 0.274 |
| **$R_1$** | 0.1561 |
| **$wR_2$** | 0.3589 |
| **Goodness of fit on $F^2$** | 0.979 |
| **$\Delta\rho_{max}$($e$ / Å$^3$)** | 0.925 |
| **$\Delta\rho_{min}$($e$ / Å$^3$)** | -0.817 |
| **Data collection** | APS, 16-IDB beamline, Pilatus Si 1M detector, λ = 0.34453 Å |

| **Atomic coordinates and equivalent displacement parameters $U_{eq}$*** | | | | | | |
|---|---|---|---|---|---|---|
| **Atom** | **Wyckoff position** | ***x*** | ***y*** | ***z*** | **$U_{eq}$ (Å$^2$)** | |
| S1 | 8*f* | 0.1540(7) | 0.0826(4) | 0.4448(5) | 0.020(4) | |
| S2 | 8*f* | 0.1957(7) | 0.0608(4) | 0.7222(5) | 0.024(4) | |
| S3 | 8*f* | 0.0309(6) | 0.3899(4) | 0.8883(5) | 0.017(4) | |

| **Anisotropic displacement parameters, in Å$^2$** | | | | | | |
|---|---|---|---|---|---|---|
| **Atom** | **$U_{11}$** | **$U_{22}$** | **$U_{33}$** | **$U_{12}$** | **$U_{13}$** | **$U_{23}$** |
| S1 | 0.016(7) | 0.0094(11) | 0.041(8) | -0.0015(15) | 0.021(7) | -0.0002(16) |
| S2 | 0.020(7) | 0.0082(12) | 0.051(9) | -0.0003(14) | 0.029(8) | 0.0004(15) |
| S3 | 0.010(7) | 0.0092(13) | 0.037(9) | 0.0006(12) | 0.018(8) | 0.0007(14) |

*$U_{eq}$ is defined as one third of the trace of the orthogonalized $U^{ij}$ tensor.

## Table S2. Details of crystal structure refinement for *Pnma* phase at 132 GPa

| **Sample name** | **AG016_p01_q1_s2** | | | | |
|---|---|---|---|---|---|
| **Pressure, GPa** | 132(2) | | | | |
| **Temperature of annealing, K** | 1500(100) | | | | |
| ***a*** **(Å)** | 7.631(3) | | | | |
| ***b*** **(Å)** | 4.469(6) | | | | |
| ***c*** **(Å)** | 7.6906(19) | | | | |
| ***V*** **($Å^3$)** | 262.3(4) | | | | |
| ***Z*** | 8 | | | | |
| **$2\Theta_{min}$ for data collection (°)** | 2.2 | | | | |
| **$2\Theta_{max}$ for data collection (°)** | 15.032 | | | | |
| **Completeness to *d* = 0.8 Å** | 0.532 | | | | |
| **Reflections collected** | 513 | | | | |
| **Independent reflections** | 293 | | | | |
| **Independent reflections [$I > 2\sigma(I)$]** | 226 | | | | |
| **Refined parameters** | 25 | | | | |
| **$R_{int}(F^2)$** | 0.0506 | | | | |
| **$R(\sigma)$** | 0.047 | | | | |
| **$R_1$ [$I > 2\sigma(I)$]** | 0.0732 | | | | |
| **$wR_2$ [$I > 2\sigma(I)$]** | 0.1926 | | | | |
| **$R_1$** | 0.0905 | | | | |
| **$wR_2$** | 0.209 | | | | |
| **Goodness of fit on $F^2$** | 1.138 | | | | |
| **$\Delta\rho_{max}(e / Å^3)$** | 1.087 | | | | |
| **$\Delta\rho_{min}(e / Å^3)$** | -1.138 | | | | |
| **Data collection** | APS, 13-IDD beamline, Pilatus CdTe 1M detector, λ = 0.29520 Å | | | | |

| **Atomic coordinates and equivalent displacement parameters $U_{eq}$*** | | | | | | |
|---|---|---|---|---|---|---|
| **Atom** | **Wyckoff position** | ***x*** | ***y*** | ***z*** | **$U_{eq}$ ($Å^2$)** | |
| S1 | 4*c* | 0.2358(3) | 0.75 | 0.1520(3) | 0.0054(11) | |
| S2 | 4*c* | 0.0737(3) | 0.75 | 0.4929(2) | 0.0057(11) | |
| S3 | 4*c* | 0.2394(3) | 0.75 | 0.8137(3) | 0.0059(10) | |
| S4 | 4*c* | 0.0599(3) | 0.25 | 0.9876(2) | 0.0045(10) | |

| **Anisotropic displacement parameters, in $Å^2$** | | | | | | |
|---|---|---|---|---|---|---|
| **Atom** | **$U_{11}$** | **$U_{22}$** | **$U_{33}$** | **$U_{23}$** | **$U_{13}$** | **$U_{12}$** |
| S1 | 0.0084(11) | 0.001(4) | 0.0071(10) | 0 | 0.0010(5) | 0 |
| S2 | 0.0059(10) | 0.005(3) | 0.0065(9) | 0 | -0.0012(5) | 0 |
| S3 | 0.0088(11) | 0.003(3) | 0.0057(9) | 0 | 0.0018(5) | 0 |
| S4 | 0.0051(12) | 0.003(3) | 0.0051(9) | 0 | 0.0018(4) | 0 |

*$U_{eq}$ is defined as one third of the trace of the orthogonalized $U^{ij}$ tensor.

## Table S3. Details of crystal structure refinement for *Pnma* phase at 157 GPa

| | | | | | | |
|---|---|---|---|---|---|---|
| **Sample name** | **AG016_p02_s1** | | | | | |
| **Pressure, GPa** | **157(2)** | | | | | |
| **Temperature of annealing, K** | **not annealed** | | | | | |
| ***a*** **(Å)** | 7.469(6) | | | | | |
| ***b*** **(Å)** | 4.3880(18) | | | | | |
| ***c*** **(Å)** | 7.574(2) | | | | | |
| $V$ **(Å³)** | 248.2(2) | | | | | |
| $Z$ | 8 | | | | | |
| **$2\Theta_{min}$ for data collection (°)** | 2.228 | | | | | |
| **$2\Theta_{max}$ for data collection (°)** | 15.169 | | | | | |
| **Completeness to $d$ = 0.8 Å** | 0.586 | | | | | |
| **Reflections collected** | 483 | | | | | |
| **Independent reflections** | 319 | | | | | |
| **Independent reflections [$I > 2\sigma(I)$]** | 238 | | | | | |
| **Refined parameters** | 25 | | | | | |
| $R_{int}(F^2)$ | 0.0168 | | | | | |
| $R(\sigma)$ | 0.0299 | | | | | |
| $R_1$ [$I > 2\sigma(I)$] | 0.0666 | | | | | |
| $wR_2$ [$I > 2\sigma(I)$] | 0.1818 | | | | | |
| $R_1$ | 0.0822 | | | | | |
| $wR_2$ | 0.1975 | | | | | |
| **Goodness of fit on $F^2$** | 1.089 | | | | | |
| $\Delta\rho_{max}(e / Å^3)$ | 1.341 | | | | | |
| $\Delta\rho_{min}(e / Å^3)$ | -1.331 | | | | | |
| **Data collection** | APS, 13-IDD beamline, Pilatus CdTe 1M detector, λ = 0.29520 Å | | | | | |
| **Atomic coordinates and equivalent displacement parameters $U_{eq}$*** | | | | | | |
| **Atom** | **Wyckoff position** | ***x*** | ***y*** | ***z*** | $U_{eq}$ **(Å²)** | |
| S1 | 4*c* | 0.2346(3) | 0.75 | 0.1514(2) | 0.0051(6) | |
| S2 | 4*c* | 0.0724(3) | 0.75 | 0.4925(2) | 0.0044(5) | |
| S3 | 4*c* | 0.2376(3) | 0.75 | 0.8113(2) | 0.0048(5) | |
| S4 | 4*c* | 0.0570(3) | 0.25 | 0.9865(2) | 0.0044(5) | |
| **Anisotropic displacement parameters, in Å²** | | | | | | |
| **Atom** | $U_{11}$ | $U_{22}$ | $U_{33}$ | $U_{23}$ | $U_{13}$ | $U_{12}$ |
| S1 | 0.0072(16) | 0.0018(9) | 0.0062(9) | 0 | 0.0002(5) | 0 |
| S2 | 0.0036(15) | 0.0021(10) | 0.0074(9) | 0 | -0.0004(5) | 0 |
| S3 | 0.0074(15) | 0.0014(9) | 0.0056(9) | 0 | 0.0011(5) | 0 |
| S4 | 0.0085(15) | 0.0008(9) | 0.0039(8) | 0 | 0.0006(5) | 0 |

*$U_{eq}$ is defined as one third of the trace of the orthogonalized $U^{ij}$ tensor.

## Table S4. Details of crystal structure refinements for *Im*-3*m* phase at high pressures*

| **Sample name** | **AG016_ p02_q1_s1** | **AG016_ p03_q1_s1** | **AG016_ p04_s1** | **AG016_ p05_s1** | **AG016_ p06_q1_s1** |
|---|---|---|---|---|---|
| **Pressure, GPa** | **159(2)** | **190(2)** | **204(2)** | **219(2)** | **224(2)** |
| **Temperature of annealing, K** | **1600(100)** | **1700(100)** | **not annealed** | **not annealed** | **1800(100)** |
| $a$ (Å) | 3.0938(6) | 3.0657(7) | 3.019(2) | 2.9983(17) | 3.0120(9) |
| $V$ (Å$^3$) | 29.611(19) | 28.81(2) | 27.51(6) | 26.96(5) | 27.33(2) |
| **$2\Theta_{min}$ for data collection (°)** | 3.869 | 3.904 | 3.965 | 3.992 | 3.974 |
| **$2\Theta_{max}$ for data collection (°)** | 15.148 | 14.211 | 12.631 | 12.718 | 14.47 |
| **Completeness to $d$ = 0.8 Å** | 0.857 | 0.857 | 0.833 | 0.833 | 1 |
| **Reflections collected** | 23 | 29 | 24 | 27 | 28 |
| **Independent reflections** | 10 | 11 | 9 | 9 | 9 |
| **Independent reflections [$I > 2\sigma(I)$]** | 10 | 11 | 9 | 9 | 9 |
| **Refined parameters** | 2 | 2 | 2 | 2 | 2 |
| $R_{int}(F^2)$ | 0.0584 | 0.0691 | 0.0525 | 0.0402 | 0.0855 |
| $R(\sigma)$ | 0.035 | 0.0544 | 0.025 | 0.0185 | 0.0473 |
| $R_1$ [$I > 2\sigma(I)$] | 0.0349 | 0.0314 | 0.0379 | 0.0407 | 0.0216 |
| $wR_2$ [$I > 2\sigma(I)$] | 0.0834 | 0.0527 | 0.1036 | 0.1024 | 0.0364 |
| $R_1$ | 0.0349 | 0.0314 | 0.0379 | 0.0407 | 0.0216 |
| $wR_2$ | 0.0834 | 0.0527 | 0.1036 | 0.1024 | 0.0364 |
| **Goodness of fit on $F^2$** | 1.208 | 1.172 | 1.364 | 1.228 | 1.04 |
| $\Delta\rho_{max}$($e$ / Å$^3$) | 0.413 | 0.437 | -0.468 | 0.684 | 0.514 |
| $\Delta\rho_{min}$($e$ / Å$^3$) | -0.549 | -0.5 | -0.468 | -0.402 | -0.31 |
| $U_{iso}$(S1) (Å$^2$) | 0.0051(8) | 0.0069(5) | 0.0083(16) | 0.0117(18) | 0.0095(6) |
| **Data collection** | APS, 13-IDD beamline, Pilatus CdTe 1M detector, λ = 0.29520 Å | | | | |

*$H_3$(S,C) crystallizes in space group *Im*-3*m*, $Z$ = 2, atoms' Wyckoff positions:
S1 2*a* (0, 0, 0)
H1 6*b* (0, 0, 0.5)

## References

[1] V.B. Prakapenka, A. Kubo, A. Kuznetsov, A. Laskin, O. Shkurikhin, P. Dera, M.L. Rivers, and S.R. Sutton, High Press. Res. **28**, 225 (2008).

[2] C. Prescher and V.B. Prakapenka, High Press. Res. **35**, 223 (2015).

[3] CrysAlisPro Softw. Syst. Version 1.171.40.84a, Rigaku Oxford Diffraction, Oxford, UK (2019).

[4] E. Bykova, M. Bykov, S. Chariton, V.B. Prakapenka, K. Glazyrin, A. Aslandukov, A. Aslandukova, G. Criniti, A. Kurnosov, and A.F. Goncharov, Phys. Rev. B **103**, L140105 (2021).

[5] D. Duan, Y. Liu, F. Tian, D. Li, X. Huang, Z. Zhao, H. Yu, B. Liu, W. Tian, and T. Cui, Sci. Rep. **4**, 30 (2014).

[6] G.M. Sheldrick, Acta Crystallogr. Sect. C Struct. Chem. **71**, 3 (2015).

[7] O. V. Dolomanov, L.J. Bourhis, R.J. Gildea, J.A.K. Howard, and H. Puschmann, J. Appl. Crystallogr. **42**, 339 (2009).